\documentclass[a4paper,twocolumn]{article} 

\usepackage{libertine}

\usepackage{graphicx,relsize,afterpage,amsmath,amssymb,bm,textcomp,microtype}
\usepackage[usenames,dvipsnames,svgnames,table]{xcolor}
\usepackage{soul,sfmath,authblk,mathtools,float,gensymb}
\usepackage[top=2cm,bottom=2cm,left=1.5cm,right=1.5cm]{geometry}
\usepackage{scicite}
\usepackage{balance}

\usepackage{lineno}
\usepackage{comment}
\usepackage[misc]{ifsym}
\usepackage[hyperfootnotes=false,colorlinks=false,linkcolor=black,citecolor=black,urlcolor=black,draft]{hyperref}
\usepackage[font=small,labelfont=bf]{caption}
\DeclareCaptionLabelSeparator{lsep}{ $|$ } 
\newcommand\blfootnote[1]{%
  \begingroup
  \renewcommand\thefootnote{}\footnote{#1}%
  \addtocounter{footnote}{-1}%
  \endgroup
}

\makeatletter
\renewenvironment{abstract}{%
    \if@twocolumn
      \section*{\abstractname}%
    \else 
      \quotation \noindent{\bfseries \abstractname:}
    \fi}
    {\if@twocolumn\else\endquotation\fi}
\makeatother

\usepackage[nameinlink]{cleveref}
\graphicspath{{./Images/}}

\newcommand{\de}{\mathrm{d}}
\newcommand{\R}{\mathbb{R}}

\newcommand{\cC}{\mathcal{C}}
\newcommand{\cI}{\mathcal{I}}
\newcommand{\cR}{\mathcal{R}}
\newcommand{\cN}{\mathcal{N}}
\newcommand{\cO}{\mathcal{O}}
\newcommand{\cW}{\mathcal{W}}

\DeclareMathOperator{\tr}{tr}

\newcommand{\methods}{Materials and Methods}

\newcommand{\suppl}{Supplementary Note}

\creflabelformat{equation}{#2#1#3}

 \makeatletter
\newcommand\notsotiny{\@setfontsize\notsotiny\@vipt\@viipt}
\makeatother

\title{\bf Efficient Communication over Complex Dynamical Networks: The Role of Matrix Non-Normality}

\author[1]{Giacomo Baggio}
\author[2,3]{Virginia Rutten}
\author[$\dagger$4]{Guillaume Hennequin}
\author[$\ast \dagger$1]{Sandro Zampieri}

\date{}

\affil[1]{{\small Department of Information Engineering, University of Padova, via Gradenigo, 6/B I-35131 Padova, Italy}}
\affil[2]{{\small Gatsby Computational Neuroscience Unit, University College London, London W1T\,4JG, UK}}
\affil[3]{{\small Janelia Research Campus, Howard Hughes Medical Institute, Ashburn, VA, USA}}
\affil[4]{{\small Computational and Biological Learning Lab, Department of Engineering, University of Cambridge, Cambridge CB2\,1PZ, UK}}

\begin{document}


\twocolumn[
  \begin{@twocolumnfalse}
\maketitle

\vspace{0.1cm}

\begin{quotation}
\noindent\textbf{One sentence summary:} Non-normal networks spread information efficiently in the face of interference between consecutive transmissions and readout noise.
\end{quotation}

\begin{abstract}
In both natural and engineered systems, communication often occurs dynamically over networks ranging from highly structured grids to largely disordered graphs.
To use, or comprehend the use of, networks as efficient communication media requires understanding of how they propagate and transform information in the face of noise.
Here, we develop a framework that enables us to examine how network structure, noise, and interference between consecutive packets jointly determine transmission performance in networks with linear dynamics at single nodes and arbitrary topologies.
Mathematically normal networks, which can be decomposed into separate low-dimensional information channels, suffer greatly from readout and interference noise. Interestingly, most details of their wiring have no impact on transmission quality.
Non-normal networks, however, can largely cancel the effect of noise by transiently amplifying select input dimensions while ignoring others, resulting in higher net information throughput.
Our theory could inform the design of new communication networks, as well as the optimal use of existing ones.
\end{abstract}
\vspace{0.5cm}
\end{@twocolumnfalse}
]

\begin{figure*}
\includegraphics[scale=0.825]{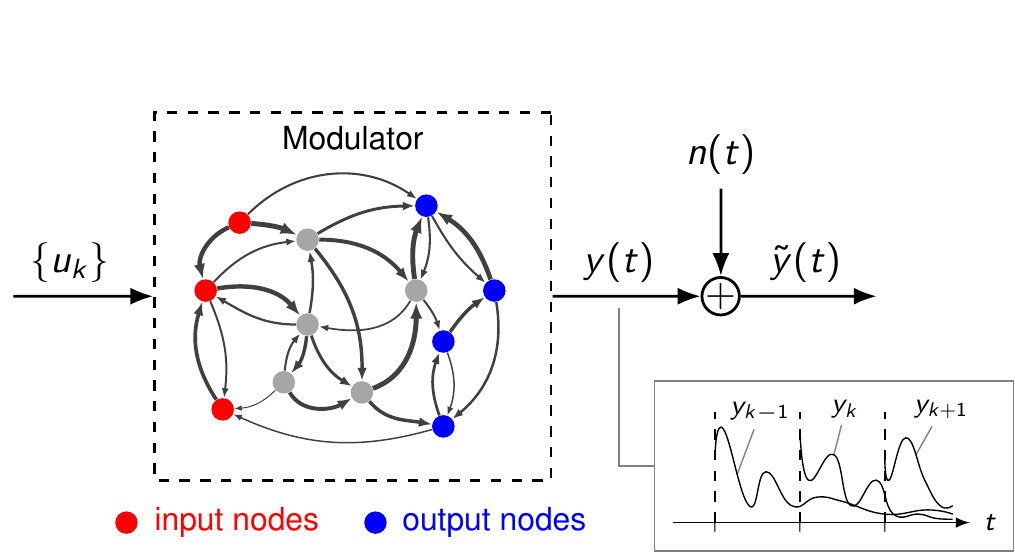}
\par
 
\vspace*{\dimexpr-\parskip-137.5pt\relax}
  \parshape 20 
    .525\textwidth .475\textwidth 
    .525\textwidth .475\textwidth 
    .525\textwidth .475\textwidth
    .525\textwidth .475\textwidth
    .525\textwidth .475\textwidth
    .525\textwidth .475\textwidth
    .525\textwidth .475\textwidth
    .525\textwidth .475\textwidth
    .525\textwidth .475\textwidth
    .525\textwidth .475\textwidth
    .525\textwidth .475\textwidth
    .525\textwidth .475\textwidth
    .525\textwidth .475\textwidth
    .525\textwidth .475\textwidth
    .525\textwidth .475\textwidth
    .525\textwidth .475\textwidth
    .525\textwidth .475\textwidth
    .525\textwidth .475\textwidth
    .525\textwidth .475\textwidth
    0pt \textwidth 
  \makeatletter
  \refstepcounter\@captype
  \addcontentsline{\csname ext@\@captype\endcsname}{\@captype}
    {\protect\numberline{\csname the\@captype\endcsname}{ToC entry}}%
  {\small \bfseries \csname fnum@\@captype\endcsname\ $|$ }
  \makeatother
  \small {  \textbf{Channel description.}
  A sequence of to-be-transmitted packets of information is encoded in a sequence of random vectors $u_k\in\R^{m}$ independently drawn from an identical encoding probability distribution $p(u)$. Each $u_k$ excites the input nodes (red) of a complex dynamical network (\Cref{eq:linearsys1}).
The dynamics of this network act as a ``modulator'', producing activation trajectories \smash{$y(t)=\sum_{k}y_{k}(t)$}, with \smash{$y_{k}(t)=Ce^{A(t-kT)}Bu_{k}\mathbf{1}(t-kT)$} being the ``modulated'' waveform corresponding to the input vector $u_k$, in some output nodes (blue).} These are further corrupted by independent Gaussian noise $n(t)$ before reaching the receiver.
Importantly, the modulator is not memoryless. In fact, due to first-order dynamics in each node of the network, patterns of network activity elicited by previously transmitted packets linger and interfere with the current communication, thus effectively contributing an intrinsic source of structured noise that adds up to the readout noise.
\label{Fig:schematics}
\vspace{0.1cm}
\end{figure*}

\section*{ Introduction}

Reliable propagation of information through networks with unreliable nodes is a fundamental problem facing many engineered and natural systems. This includes social networks \cite{Guille13,Molaei18}, peer-to-peer networks \cite{Decker13}, gene regulatory networks \cite{Cheong11,Selimkhanov14}, power grids \cite{Galli11}, and brain networks \cite{Laughlin03, Avena18}, to cite only a few.\blfootnote{\textsuperscript{$\ast$}Corresponding author: \href{mailto:sandro.zampieri@unipd.it}{sandro.zampieri@unipd.it}.}\blfootnote{\textsuperscript{$\dagger$}Equal contributions.}
%
In order to engineer better communication networks, make better use of existing ones, or understand how natural (e.g., biological) networked systems function, a theory is needed that relates the network's connectivity and dynamics to its performance in transmitting information.

Previous work at the interface of network science and information theory has been largely restricted to static, feedforward networks, in which packets of activity travel one after the other through layers of memoryless nodes, with no interference.
Examples include classic connectionist work where feedforward ``neural'' networks are optimized so their outputs retain as much information as possible about their inputs \cite{Toyoizumi05,Sharpee07}.
These works have influenced how neuroscientists think about sensory pathways, which resemble layered networks of noisy neurons receiving input packets from body senses \cite{Kandel00}.
In particular, the neural representations of visual stimuli that are found along the primate ventral stream are strikingly similar to those that emerge in deep networks trained on object recognition tasks \cite{Yamins16}.
More recent work \cite{Shwartz17} has drawn a link between deep learning \cite{Goodfellow16} and the information bottleneck method \cite{Tishby00}, a principled approach to compressive communication.
Beyond feedforward networks, the effect of recurrent topologies on information transmission was studied in the context of virtual electrical circuits \cite{Rubido17}, but this was restricted to steady states and therefore disregarded any potential encoding of information in activity transients.

In most real-world scenarios, however, information does not propagate statically (or instantaneously), but dynamically within complex recurrent networks composed of non-memoryless nodes.
The inherent dynamics of the network can greatly affect communication performance in ways that remain poorly understood.
In \cite{Kirst16}, the authors proposed an analytical framework
based atop standard notions of time-delayed mutual information and transfer entropy,
to quantify the routing of small activity fluctuations propagating on top of oscillatory reference dynamics.
While their framework allowed them to identify a generic mechanism capable of generating flexible information-routing patterns in the network, it is based on a small-noise approximation and therefore cannot fully capture the impact of noise on network communication.
Moreover, the authors did not systematically study the role of network topology.  
The authors of \cite{Harush17} investigated the interplay between the network topology and its dynamics.
They found that patterns of information are governed by universal laws that depend only on a few relevant parameters of the network dynamics.
However, the analysis was carried out in a deterministic setting, and the proposed information transfer metric | which quantifies the sensitivity of a dynamical system to local perturbations | lacks an explicit information-theoretic interpretation.
The work \cite{Ganguli08} used Fisher information theory to quantify the short-term memory storage capacity of networks governed by linear dynamics.
In investigating this memory problem, which is a form of network communication through time, the authors were led to study the interactions between single-node dynamics, connectivity, and input statistics, similar to the theory we develop here.
However, the network received a one-dimensional input, and temporal correlations~were~neglected. 


Here, we study the role of graph topology on the quality of information transmission in noisy networks with otherwise simple, linear single-node dynamics.
We establish a novel framework for quantifying the maximum amount of information about high-dimensional inputs that can be transmitted reliably through such networks. 
We apply our framework to various network architectures, ranging from simple, structured networks amenable to analytical derivations, to more complex, disordered, and real networks that we investigate numerically.
Critically, all the networks we consider here have memory, from which interference arises between the network's response to multiple packets transmitted in close succession, and constitutes a source of internal, structured noise.
We show that when the amount of noise present in the information channel is large, anisotropic (mathematically ``non-normal'', \cite{Trefethen2005,Asllani18}) networks that embed directed feedforward pathways perform better than isotropic (``normal'') ones.\footnote{  A network is said to be normal if its connectivity matrix $A$ is normal, i.e., if it satifies $AA^\dagger = A^\dagger A$, where $\cdot^\dagger$ denotes the conjugate transpose. Otherwise, the network is said to be non-normal  \cite{Asllani18}.} 
Moreover, we find that such non-normal networks can even entirely overcome the effect of noise in some limit.
Our results provide estimates for the amount of information that a network can propagate, and insights into how the propagation of information depends on key network properties. 
Additionally, we discuss how information propagation can be optimized by using specific distributions of input packets.
We expect our theory to contribute to understanding the behaviour of natural networked systems, which are often found to be strongly non-normal \cite{ALC18}.
Further dissection of the mechanisms at work in natural networks (e.g., single-node dynamics, graph structure, adaptive wiring, \ldots) may also suggest better engineered solutions to network communication.

\section*{Results}

\subsection*{Modelling framework}

\subsubsection*{Communication through networks}
We consider the following model of a communication channel, whereby a sequence of to-be-transmitted packets of information is probabilistically encoded in a sequence of input vectors (\Cref{Fig:schematics}). Information transmission occurs via propagation of the inputs through a dynamical network. In order to obtain analytical, interpretable results that hold for arbitrarily complex graph topologies, we assume minimalistic dynamics for single network nodes: first-order, linear responses to inputs. Specifically, we consider continuous-time linear dynamical systems of the form
\begin{align}\label{eq:linearsys1}
\begin{split}
	\frac{\de x(t)}{\de t} &= A x(t) + B\, {  \sum_{k} u_{k}\, \delta(t-kT)}, \\
	y(t) &= C x(t), 
\end{split}
\end{align}
where $x(t)\in\R^{n}$ denotes the state vector and $A\in\R^{n\times n}$ is the state matrix.
We restrict our analysis to the case of ``stable'' network dynamics, whereby responses to transient inputs do not grow unbounded (which would be physically unfeasible) but fade away after some time.
Mathematically, this means we require all eigenvalues of $A$ to have negative real part.

Each input vector $u_k\in\R^{m}$, {  independently drawn from an identical encoding probability distribution $p(u)$, contains the information carried by the $k$-th transmitted packet.}
Each of these inputs is then delivered as an impulse (here modelled as a Dirac's delta $\delta(\cdot)$)  that excites the network dynamics in \Cref{eq:linearsys1}.
{  Transmission of successive packets occurs every $T$ units of time.}
The columns of the matrix $B\in\R^{n\times m}$ define ``input nodes'' (red circles in \Cref{Fig:schematics}), which are the only ones affected by the impulse.
Likewise, a readout matrix $C\in\R^{p\times n}$ singles out specific output nodes (blue circles) whose activations $y(t)$ are transmitted to the receiver, further corrupted by independent Gaussian noise of variance $\sigma^2$. {  This results in corrupted trajectories $\tilde{y}(t)$ which the receiver could use to reconstruct the corresponding input packets. 
In our assessment of communication performance, we will consider Shannon's mutual information as a proxy for reconstruction quality (see below), instead of considering explicit decoding~algorithms.
}

By reducing the complexity of single node dynamics to simple first-order evolution, \Cref{eq:linearsys1} allows us to focus on the effect of network architecture 
on the quality of information transmission.
For example, \Cref{eq:linearsys1} is known as a ``rate equation'' in computational neuroscience, whereby it has been shown to capture key aspects of the dynamics of neuronal networks around fixed points \cite{DA01}.
Indeed, single neurons are often characterized by input/output functions that remain approximately linear over their relevant dynamic range \cite{Inagaki17}. In that case, $A$ represents the matrix of synaptic connection weights, and $x(t)$ is interpreted as momentary deviations from steady-state firing~rates.

Importantly, since each network node is governed by first-order dynamics, the network is not memoryless: activity trajectories elicited by previous communications interfere with (in fact, add linearly to) the network trajectory carrying information about the current input. Thus, for the transmission of a packet at time $t=0$ (assuming many packets have already been transmitted), interference contributes an additional source of noise $i(t)$, given by
\begin{align}
    i(t) =  \sum_{k=1}^{\infty} C e^{A(t+k T)} B\, u_{-k}{ \mathbf{1}(t+kT)},
\end{align}
where $\mathbf{1}(\cdot)$ denotes the unit-step function, defined as $\mathbf{1}(x)=0$, if $x<0$, and $\mathbf{1}(x)=1$, otherwise.
This phenomenon, known as inter-symbol interference in communications \cite{Sklar01}, arises in any communication medium that has some form of memory, including networks with node dynamics described by differential equations. 

In the following, we study the combined effects of the network architecture (matrix $A$), communication time window ($T$), noise level ($\sigma^2$), and encoding of input packets under this communication paradigm.
We begin by establishing an analytical framework to characterize the quality of information transmission through the network, and highlight the trade-off that arises between sending packets of information at a high temporal rate and the ability for the receiver to accurately reconstruct them. We then summarize our analytical results, and illustrate them using appropriate network architectures.
\begin{figure*}
\includegraphics[scale=0.8]{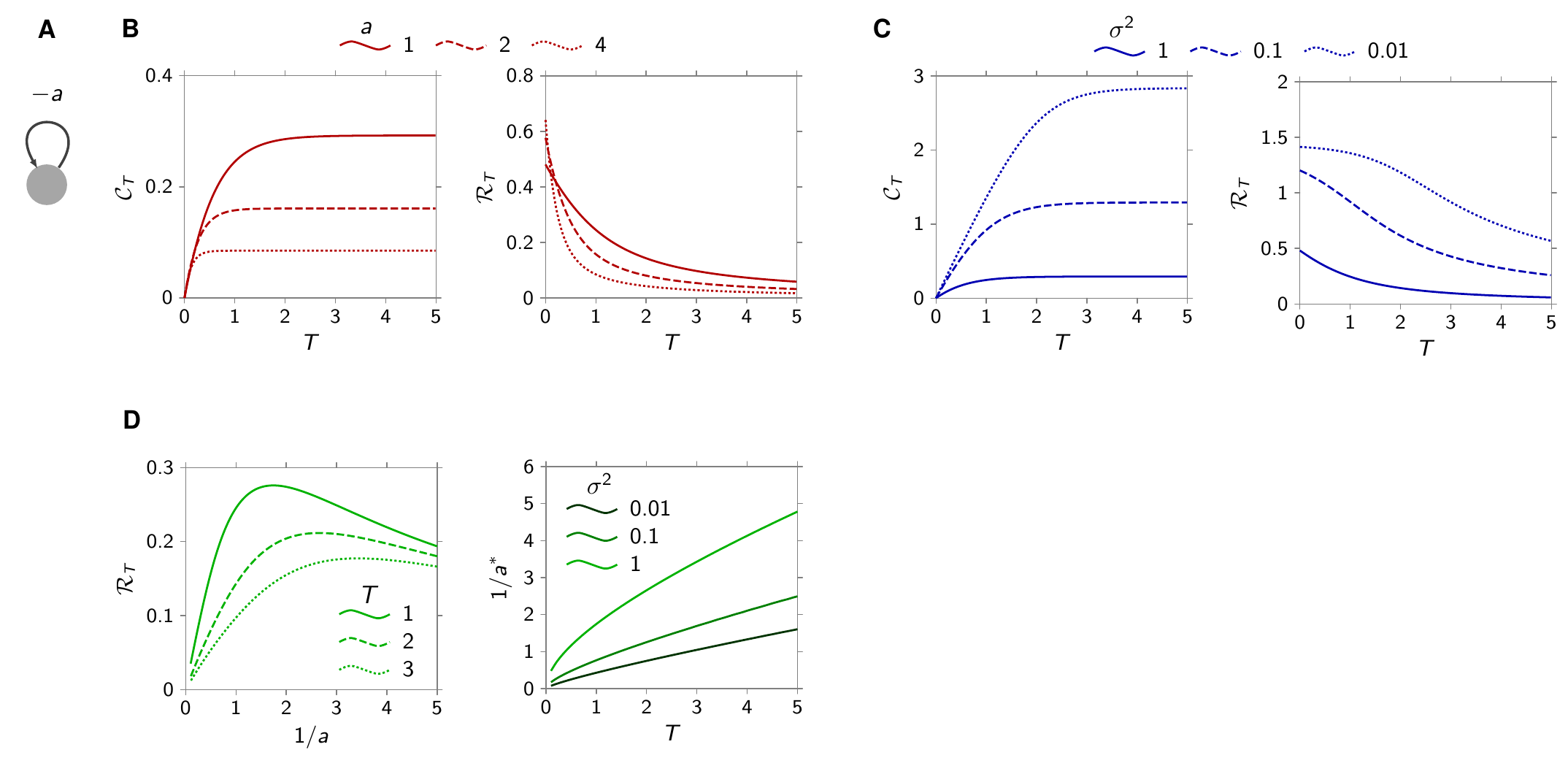}
\par
 
\vspace*{\dimexpr-\parskip-100pt\relax}
  \parshape 25 
    .55\textwidth .45\textwidth 
    .55\textwidth .45\textwidth 
    .55\textwidth .45\textwidth
    .55\textwidth .45\textwidth
    .55\textwidth .45\textwidth
    .55\textwidth .45\textwidth
    .55\textwidth .45\textwidth
    .55\textwidth .45\textwidth
    .55\textwidth .45\textwidth
    .55\textwidth .45\textwidth
    .55\textwidth .45\textwidth
    .55\textwidth .45\textwidth
    .55\textwidth .45\textwidth
    .55\textwidth .45\textwidth
    .55\textwidth .45\textwidth
    .55\textwidth .45\textwidth
    .55\textwidth .45\textwidth
    .55\textwidth .45\textwidth
    .55\textwidth .45\textwidth
    .55\textwidth .45\textwidth
    .55\textwidth .45\textwidth
    .55\textwidth .45\textwidth
    .55\textwidth .45\textwidth
    .55\textwidth .45\textwidth
    0pt \textwidth 
  \makeatletter
  \refstepcounter\@captype
  \addcontentsline{\csname ext@\@captype\endcsname}{\@captype}
    {\protect\numberline{\csname the\@captype\endcsname}{ToC entry}}%
  {\small \bfseries \csname fnum@\@captype\endcsname\ $|$ }
  \makeatother
  \small \textbf{Information transmission through a single node.}
  {\bfseries (A)}~Single node schematics.
  {\bfseries (B-C)}~Capacity $\cC_T$ and rate $\cR_T$ as functions of the transmission window $T$, for various values of the decay rate $a$ and fixed noise level $\sigma^2=1$ (B), and various values of $\sigma^2$ and fixed $a=1$ (C).
  {\bfseries (D)}~$\cR_T$ as a function of the decay time constant $1/a$ exhibits a maximum ($1/a^*$) for any value of $T$ (left, $\sigma^2=1$); this maximum grows with $T$ (right).
\label{Fig:scalar case}
\vspace{0.75cm}
\end{figure*}

\subsubsection*{Information transmission metrics}

{  To quantify the amount of information that can be propagated through the network channel described above, we use the notion of Shannon's mutual information between the input packet $u_k$ and the corresponding noisy network output $\tilde{y}(t)$  observed over the subsequent time interval $kT\leq t < (k+1)T$. Denoting by $\tilde{Y}_k$ this output function (on which inter-symbol interference acts as an additional source of noise), and assuming stationarity to drop the $k$ subscripts, we can write the mutual information (in bits) between $u$ and $\tilde{Y}$ as
\begin{align}
\label{eq:information}
\cI_T(u, \tilde{Y}) = & \int  p(u)\, \text{d}u \int p_T(\tilde{Y}|u)\log_2 \frac{p_T(\tilde{Y}|u)}{p_T(\tilde{Y})}\, \text{d}\tilde{Y},
\end{align}
where the $\cdot_T$ notation emphasises the dependence of mutual information on the transmission window
(a more formal definition of the integral over functions $\tilde{Y}$ in \Cref{eq:information} is given in \suppl\ 2). 
To better utilise the channel, the sender can use the encoding distribution $p(u)$ that maximizes the mutual information; this optimum defines an information metric which is independent of the encoding distribution,
\begin{equation}
    \label{eq:capacity_def}
    \cC_T = \max_{p(u)}\, \cI_T(u,\tilde{Y}).
\end{equation}
With a slight abuse of terminology, we will refer to this metric as information capacity, or, simply, capacity.\footnote{  Our choice of terminology is motivated by the fact that this metric coincides with the standard capacity of a digital communication channel, when the channel is memoryless, see, e.g., \cite{CT12}. We refer to Supplementary Note\ 2  for further details on the relation between the channel capacity and our metric in \Cref{eq:capacity_def}.}}

In \Cref{eq:capacity_def}, the maximization over the encoding distribution $p(u)$ must be performed with an additional constraint on input power {  (input covariance).}
Theoretically, this is required so that the capacity remains finite (the signal-to-noise ratio can be made arbitrarily large if inputs can be arbitrarily large too).
In practice, the nodes of any physical network have limited dynamic range, and therefore network inputs must be power-limited.
{  Here, we consider Gaussian encoding distributions with zero mean and covariance $\Sigma\succcurlyeq 0$, and input power constraint of the form $\text{tr}(\Sigma) \le 1$ (without loss of generality; cf. \suppl\ 3).}

\subsubsection*{An expression for the {  information capacity}}

Our main theoretical result is the following expression for the {  information capacity} (\suppl\ 2):
\begin{align}
  \cC_T = \frac{1}{2} \max_{\Sigma\succcurlyeq 0, \tr\Sigma=1} \log_2 \frac{\det \left(\sigma^2 I+ \cO \cW \right)}{\det \left(\sigma^2 I+ \cO (\cW-B\Sigma B^{\top}) \right)}, \label{eq:cap}
\end{align}
where {  $\sigma^2$ is the variance of the noise at the receiver,}  $\cO$ denotes the observability Gramian over the interval $[0,T]$ of the system in \Cref{eq:linearsys1}, and $\cW$ is the infinite-horizon controllability Gramian of the dynamics in \Cref{eq:linearsys1} discretized with sampling time $T$ and input matrix $B\Sigma^{1/2}$ \cite{H09}. 
The formal definition of these matrices is reported in \methods\ and their properties discussed in our \suppl\ 1.
Note that \Cref{eq:cap} still involves a (difficult) maximization over the input distribution (via its covariance matrix $\Sigma$); in the following, we perform this optimization analytically where possible, but otherwise numerically using efficient algorithms (\methods).

The {  information capacity} affords a few intuitive properties (cf. \suppl\ 3).
First, $\cC_T$ always grows with increasing SNR $=1/\sigma^2$.
Second, $\cC_T$ is a bounded function of $T$ that attains its maximum as $T$ grows to infinity. This is because, for increasing $T$, (i) network activations left over from previous transmissions have more time to decay away, leading to weaker interference, and (ii) longer stretches of signal are available for decoding, allowing for better estimation of the input signal via additional filtering/de-noising.
Third, $\cC_T$ cannot decrease if nodes are added to either the set of input nodes, or the set of output nodes.

We also note that, in our framework, propagation of information through the network occurs over a finite time window $T$, and packets of information can only be transmitted one at a time.
Thus, a more relevant measure of information transmission performance is number of bits of information about {  $u$ contained in $\tilde{Y}$} \emph{per unit time}, i.e.,
\begin{equation}
    \label{eq:rate}
    \cR_T = \frac1T \cC_T.
\end{equation}
We term this metric information rate.
Since the information capacity is bounded (due to output noise and inter-symbol interference), $\cR_T$ always decreases with $T$ for large enough $T$.
However, we will see that there often exists a non-zero optimal transmission window $T$, at which $\cR_T$ reaches a maximum.

\subsection*{The limitations of normal networks}

As we will see later, many high-dimensional networks can be conveniently decomposed as a set of parallel, independent communication channels each transmitting information about a one-dimensional, scalar quantity.
We therefore begin our analysis of the role of connectivity in network communication by an in-depth look at a simple case, that of a single isolated node (\Cref{Fig:scalar case}A).
With $B=C=\Sigma=1$, and $A = -a<0$ (where $1/a>0$ is the node's decay time constant),
\Cref{eq:cap} simplifies considerably, yielding the following capacity:
\begin{align}
  \cC_T =\frac{1}{2} \log_2 \frac{2a\sigma^2+1}{2a\sigma^2+ e^{-2aT}}.
  \label{eq:capacity-single2}
\end{align}
This expression illuminates some additional properties of the information capacity and its dependence on network parameters.
To begin with, $\cC_T$ grows with the allotted transmission window $T$
(\Cref{Fig:scalar case}B and C, left).
Intuitively, this is because increasing the transmission window reduces inter-symbol interference, as the node's activity has more time to decay away before the next packet is transmitted.
However, while $\cC_T$ grows linearly with $T$ for small increasing $T$, it eventually saturates at a maximum value $\propto \log_2\left(1+\frac1{2a\sigma^2}\right)$ that grows both with the node's decay time constant ($1/a$; \Cref{Fig:scalar case}B, left) and with the SNR ($1/\sigma^2$; \Cref{Fig:scalar case}C, left).
Indeed, for large enough $T$, the output noise becomes the main factor limiting the capacity, and grows increasingly dominant during the transmission of a packet as the node's activity (the ``signal'') decays exponentially over time. Thus, increasing the observation time $T$ cannot indefinitely increase the ability of an ideal observer to reconstruct the input packet.

Next, as $T$ increases with diminishing returns on the capacity (cf.\ above), the rate (information per unit time, \Cref{eq:rate}) is bound to decrease
(\Cref{Fig:scalar case}B and C, right).
Thus, keeping the transmission window very short is the most effective way for a single node to transmit information under time pressure.
In this limit,
$\cR_{\max}=\frac{1}{\ln 2}\frac{a}{1+2a\sigma^2}$ bits/s can be transmitted.\smallskip

In practice though, transmission windows cannot be made arbitrarily small. For example, visual information conveyed to the brain via the optic nerve fluctuates on a timescale that is limited ``at the source'' by the rate at which objects move in the scene, and by the frequency and speed of saccadic eye movements which determine an effective sampling frequency \cite{Castet}.
Thus, we now assume a finite transmission window $T>0$.
In this case, there exists an optimal value of the decay time constant $1/a$ for both $\cR_T$ (\Cref{Fig:scalar case}D, left) and $\cC_T$ (not shown).
This reflects a trade-off between the noise and inter-symbol interference, mathematically evident from \Cref{eq:capacity-single2}, where $\cC_T$ can be seen to go to zero when $a$ is either very small or very large.
Intuitively, for small decay time constants $1/a$, inter-symbol interference becomes irrelevant, and the information capacity is limited by the effective signal to noise ratio $(1/a)/\sigma^2$, which in turn decreases with decreasing $1/a$.
Similarly, for long decay times (increasing $1/a$), inter-symbol interference dominates, and ruins the information capacity by letting the summed activities of many previous transmissions pollute the component relevant to the current packet. 
Thus, the rate (and capacity) is expected to achieve a maximum for some intermediate, optimal value of the decay time constant.
Numerically, we find that this optimal time constant scales near-linearly with the transmission window $T$~(\Cref{Fig:scalar case}D,~right).

\begin{figure*}[!t]
{\includegraphics[scale=0.85]{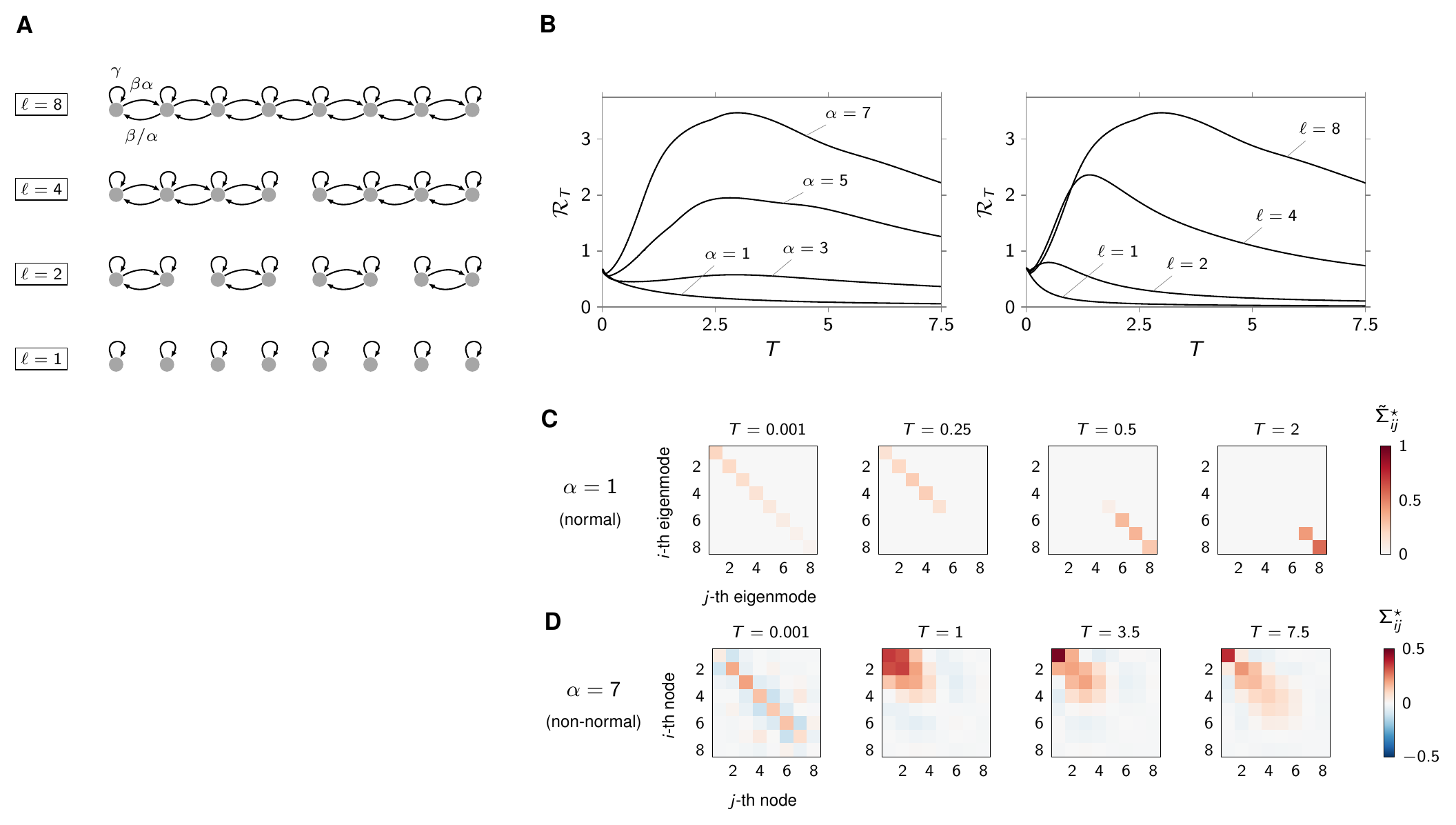}}
 \par
 
\vspace*{\dimexpr-\parskip-145pt\relax}
  \parshape 30 
    .0\textwidth .35\textwidth 
    .0\textwidth .35\textwidth 
    .0\textwidth .35\textwidth
    .0\textwidth .35\textwidth
    .0\textwidth .35\textwidth
    .0\textwidth .35\textwidth
    .0\textwidth .35\textwidth
    .0\textwidth .35\textwidth
    .0\textwidth .35\textwidth
    .0\textwidth .35\textwidth
    .0\textwidth .35\textwidth
    .0\textwidth .35\textwidth
    .0\textwidth 1.\textwidth
    .0\textwidth 1.\textwidth
    .0\textwidth 1.\textwidth
    .0\textwidth 1.\textwidth
    .0\textwidth 1.\textwidth
    .0\textwidth 1.\textwidth
    .0\textwidth 1.\textwidth
    .0\textwidth 1.\textwidth
    .0\textwidth 1.\textwidth
    .0\textwidth 1.\textwidth
    .0\textwidth 1.\textwidth
    .0\textwidth 1.\textwidth
    .0\textwidth 1.\textwidth
    .0\textwidth 1.\textwidth
    .0\textwidth 1.\textwidth
    .0\textwidth 1.\textwidth
    .0\textwidth 1.\textwidth
    0pt \textwidth 
  \makeatletter
  \refstepcounter\@captype
  \addcontentsline{\csname ext@\@captype\endcsname}{\@captype}
    {\protect\numberline{\csname the\@captype\endcsname}{ToC entry}}%
  {\small \bfseries \csname fnum@\@captype\endcsname\ $|$} 
  \makeatother
  \small \textbf{Information rate for chain network.} \textbf{(A)} Chain network schematics, with parameter values $\gamma=-2.5$, $\beta=1$, \smash{$\sigma^2=1$}, $B=I$, and $C=I$. \textbf{(B)} Plot of $\cR_T$ as a function of $T$ for four different values of $\alpha$ (for $\ell=8$), and $\ell$ (for $\alpha=7$). \textbf{(C)} Matrix plot of $\tilde\Sigma^\star$, the optimal input covariance expressed in the eigenbasis of $A$, for various values of $T$, $\ell=8$ and $\alpha=1$ (normal case). The eigenmodes of $A$ are ordered by increasing decay time constants. \textbf{(D)} Matrix plot of the optimal input covariance $\Sigma^\star$ for various values of $T$, $\ell=8$, and $\alpha=7$ (strongly non-normal~case). 
\label{Fig:line_nn_rate_full}
\vspace{0.5cm}
\end{figure*}

The case of a single-node ``network'' is, in fact, characteristic of the broader class of so-called ``normal'' networks, which include symmetric, skew-symmetric, and translation-invariant graphs to name only a few examples.
Indeed, {  when $B=C=I$},\footnote{We recall that the case $B=C=I$ represents the most favorable communication scenario, since any other choice of $B$ and $C$ provably yields a smaller value of the information capacity and rate (cf.~\suppl\ 3).} any normal network composed of $n$ nodes can be shown to behave like a set of $n$ independent scalar information channels (\suppl\ 5), each corresponding to a specific spatial ``mode'' of activity at the network level that decays at a specific rate between consecutive transmission events.
For example, for a translation-invariant architecture, these channels correspond to Fourier modes of varying spatial frequencies with decay rates that depend on the strength and spatial smoothness of the recurrent interactions \cite{benyishai1995theory,Ganguli08}.

Our mathematical analysis of normal networks shows that, despite their appealing interpretation as sets of parallel communication sub-channels, these networks might not be optimally suited for transmitting information.
First, as expected from an ensemble of independent scalar sub-channels whose rates $\cR_T$ each decrease with $T$ (recall \Cref{Fig:scalar case}B and C, right; further examples are given below), multidimensional normal networks {  with $B=C=I$} too are best exploited in the limit of very small transmission windows ($T \to 0$). 
As discussed previously, this limit is irrelevant in most applications (where $T$ is finite), implying that normal networks would always be sub-optimally exploited in practice.
Second, and more importantly, we could show that the maximum achievable performance of a normal network does not depend on the fine details of its architecture (e.g., the detailed couplings between nodes) but only on the average decay rate of its nodes (the trace of $A$).
Indeed,  for any choice of $B$ and $C$, the information rate of a normal network can never exceed (\suppl\ 5)
\begin{align}
  \cR_\text{max} = \frac1{\ln{2}} \frac{\text{tr}(A)}{2 \sigma^2 \text{tr}(A) - 1}.
  \label{eq:exact_rate_normal}
\end{align}
In particular, the above limit is attained with equality when all nodes are transmitting and receiving packets of information, that is, when $B=C=I$.
Critically, there are infinitely many network architectures that share the same $\text{tr}(A)$ but have otherwise very different geometries. {  Thus, it would be somewhat surprising if, among the very large set of all (i.e., normal and non-normal) networks with the same trace}, the restricted subset of normal networks achieved the best performance.
What is more, \Cref{eq:exact_rate_normal} also implies that the maximum rate of any normal network in the low SNR regime is simply $\cR_\text{max} \approx \frac1{2\ln{2} \sigma^2}$, which no longer depends on the connectivity matrix $A$.
In other words, no amount of clever structuring of a normal architecture can ever rescue the drop in information rate incurred by a decrease in SNR.
These considerations prompted us to study information transmission through more general, non-normal networks.

\subsection*{Role of non-normality in information transfer}

A ``non-normal'' network is any network whose connectivity matrix $A$ is not normal \cite{Trefethen2005,Asllani18}.
Thus, given the equivalence of normal networks with independent parallel channels discussed above, a non-normal network is one that cannot be so decomposed.
This implies the existence of effective feedforward pathways, embedded either explicitly at the level of network nodes (i.e., an  ``anisotropic'' tree-like structure that one would notice by looking at the connection graph; \cite{BZ18}) or implicitly at the level of orthogonal activity modes that involve many nodes simultaneously (``hidden'' feedforward pathways; \cite{Goldman09,Murphy09,Hennequin12}).
Mathematically, explicit and implicit tree-like structures can both be identified via the Schur decomposition $A = U\Delta U^\dagger$.
If $A$ is normal, this decomposition returns a diagonal matrix $\Delta$, with the Schur modes (columns of $U$) interpreted as separate information channels with decay rates given by the diagonal of $\Delta$.
For a non-normal matrix $A$, the Schur decomposition returns a triangular $\Delta$, the off-diagonal elements of which reveal hidden feedforward connections between the Schur~modes.

While it is straightforward to classify a matrix as normal or non-normal, the extent or ``degree'' to which a matrix departs from normality, and how such departure affects the dynamics of the network and communication performance, are more difficult to assess. Indeed, although several non-normality metrics of either ``dynamical'' or ``algebraic'' nature have been proposed in the literature (\suppl\ 6), there does not exist a unique scalar parameter quantifying the amount of non-normality of general matrices. To address this, we begin with a class of linear graphs whose departure from normality is parameterized by two characteristics that we can choose independently and arbitrarily: the length of the chains embedded in the graph, and the directionality of these chains (\Cref{Fig:line_nn_rate_full}A).\footnote{More generally, it can be shown that structural indicators of network non-normality are: (i) absence of cycles, (ii) low reciprocity of directed edges, and (iii) presence of hierarchical organization (see \cite{ALC18}). However, if the network is stable, the strength and length of directional paths in the networks represent effective indicators of non-normality (cf.~\suppl\ 6 and \cite{BZ18}).}   The connectivity matrix of these networks reads
\begin{align} \label{eq:chain}
    A = \begin{bmatrix}
        \gamma & \beta/\alpha & 0 & \cdots & 0 \\
         \alpha\beta & \gamma & \beta/\alpha & \ddots & \vdots \\
         0 & \alpha\beta & \ddots & \ddots & 0 \\
         \vdots & \ddots & \ddots & \ddots & \beta/\alpha \\
          0 & \cdots & 0 & \alpha\beta & \gamma \\
        \end{bmatrix}\in\mathbb{R}^{n\times n},
\end{align}
where $\alpha$, $\beta>0$, and $\gamma<-2\beta$ to enforce stability.
The simplicity of this architecture allows us to conveniently decouple the effects of (i) the eigenvalues of $A$, and (ii) its departure from normality, on the network dynamics (see below).
We show later that the insights obtained from this simple structured example topology, especially concerning the role of network non-normality, carry over to higher-dimensional and heterogeneous networks.  In particular, analogous considerations apply to the family of ``layered'' networks described in our \suppl\ 7. This class consists of networks with arbitrary ``baseline topology'' made increasingly non-normal through a process of ``directed stratification''.  In addition, for these networks, one can define parameters $\alpha$ and $\ell$ that represent the directionality strength between adjacent layers and depth of connected layers, respectively. As in the chain network \eqref{eq:chain}, these parameters regulate departure from normality. 

Mathematically \emph{normal} versions of this chain architecture are obtained either when there effectively is no chain (set of isolated nodes), or when there is no specific directionality in the connectivity ($\alpha=1$, symmetric graph).
In either case, the information rate decreases with increasing transmission window $T$ (\Cref{Fig:line_nn_rate_full}B, lowest curves), consistent with the formal theory developed above.
To understand this behaviour, and as a preliminary to our analysis of non-normal networks, we examine the optimal allocation of input power, or the spatial structure of the optimal input distribution. 
In \Cref{Fig:line_nn_rate_full}C, we plot the optimal input covariance $\Sigma^\star$
(calculated as part of deriving the capacity; recall \Cref{eq:cap}), expressed in the eigenbasis of the connectivity matrix $A$, with eigenvectors sorted by decreasing values of their decay rate. 
For long transmission windows, more of the input variance is funnelled through slow-decaying modes than through fast-decaying ones (right, $T\geq 0.5$).
This allows more of the input signal to survive the natural decay of activity in the network, thereby sustaining the signal-to-noise ratio at the receiver.
For shorter transmission windows, this strategy no longer pays off: much of what is ``signal'' for the current transmission is effectively ``noise'' for the next transmission epoch, and prolonging its decay adds further inter-symbol interference.
Accordingly, the optimal allocation strategy for short $T$ is the opposite of that for large $T$: each sub-channel is now allocated power proportional to its decay rate (\suppl\ 5).
Finally, while achieving the information capacity requires careful selection of sub-channels according to their decay rates (as just discussed), concentrating the input power on too few channels comes at a cost, as communication no longer exploits all the network's degrees of freedom.
This is best illustrated in a set of $n$ independent nodes with identical time constants, for which the best strategy is provably to give each node an equal share $P/n$ of the total available power (\suppl\ 5).
This amounts to maximizing the entropy of the input distribution.
The covariances matrices of \Cref{Fig:line_nn_rate_full}C represent the optimal way of resolving the above trade-offs, for the chain architecture considered here.

\begin{figure*}[!t]
\includegraphics[scale=0.875]{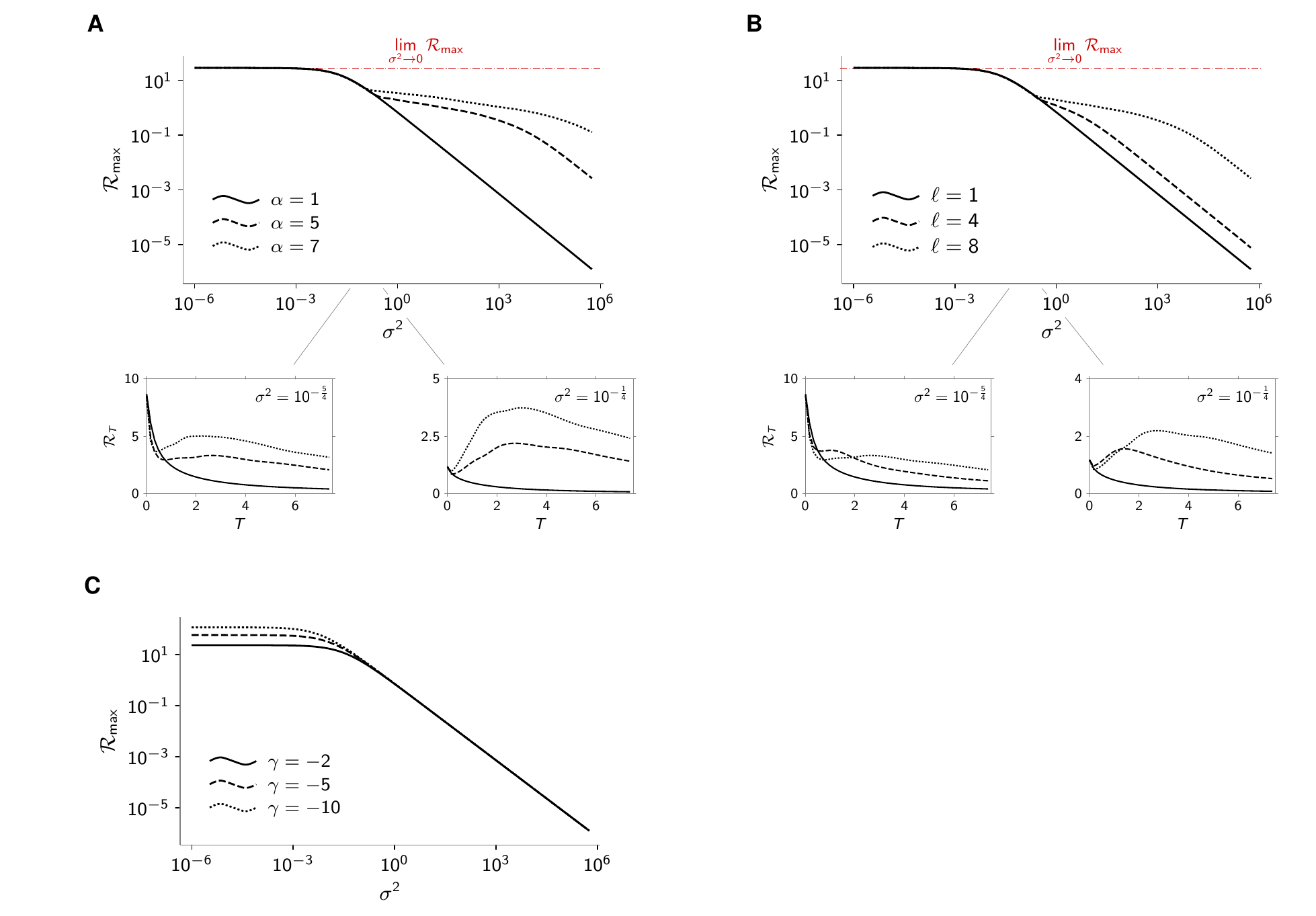} 
 \par
 
\vspace*{\dimexpr-\parskip-115pt\relax}
  \parshape 11 
    .525\textwidth .475\textwidth 
    .525\textwidth .475\textwidth 
    .525\textwidth .475\textwidth
    .525\textwidth .475\textwidth
    .525\textwidth .475\textwidth
    .525\textwidth .475\textwidth
    .525\textwidth .475\textwidth
    .525\textwidth .475\textwidth
    .525\textwidth .475\textwidth
    .525\textwidth .475\textwidth
    0pt \textwidth 
  \makeatletter
  \refstepcounter\@captype
  \addcontentsline{\csname ext@\@captype\endcsname}{\@captype}
    {\protect\numberline{\csname the\@captype\endcsname}{ToC entry}}%
  {\small \bfseries \csname fnum@\@captype\endcsname\ $|$} 
  \makeatother
  \small \textbf{Maximum information rate vs.~noise level for chain network.}  \textbf{(A)} Plot of $\cR_{\max}:=\max_{T\ge 0} \cR_{T}$ versus the output noise variance for the network of \Cref{Fig:line_nn_rate_full}A with $B=C=I$, $\gamma=-2.5$, $\ell=8$,  and three different values of $\alpha$. \textbf{(B)} Plot of $\cR_{\max}$ versus the output noise variance for the network of \Cref{Fig:line_nn_rate_full}A with $B=C=I$, $\gamma=-2.5$, $\alpha=5$, and three different values of $\ell$. \textbf{(C)} Plot of $\cR_{\max}$ versus the output noise variance for the network of \Cref{Fig:line_nn_rate_full}A with $B=C=I$, $\alpha=1$, $\ell=8$, and three different values~of~$\gamma$. \label{Fig:noise_vs_nn_line}
  \vspace{1.2cm}
\end{figure*}


We next show that large gains in information rate can be obtained by making the network connectivity non-normal. 
The degree of non-normality of the chain's connectivity matrix ($\ell=8$) can be increased, \emph{without altering its eigenvalues}, by increasing a single parameter $\alpha$ reflecting the graph's directionality (\Cref{Fig:line_nn_rate_full}A).
As the network is made increasingly non-normal in this way, its information rate grows to eventually exceed the normal networks' optimal rate by a large margin.
Moreover, the optimal rate is now attained at some realistic, finite transmission window $T$  (\Cref{Fig:line_nn_rate_full}B, left).

\begin{figure*}[t]
\begin{center}
\includegraphics[scale=0.875]{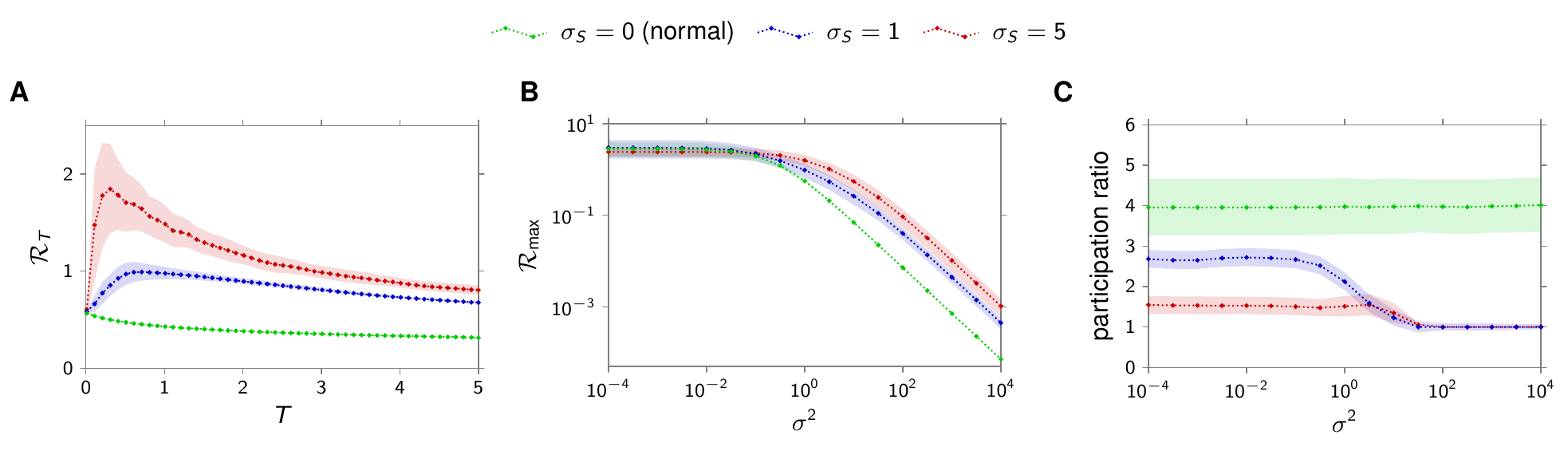} 


\end{center}
\caption{
{\bfseries Role of non-normality in high-dimensional, heterogeneous
networks.}
{\bfseries (A)}~Information rate as function of the transmission window $T$ for $n=20$, $\sigma^2=1$, $B=C=I$, and $A$ drawn according to \Cref{eq:a_gen_nn}.
   {\bfseries (B)}~Maximum information rate $\cR_{\max}:=\max_{T\ge 0} \cR_{T}$ as function of the readout noise variance $\sigma^2$.
  \textbf{(C)}~Effective dimensionality (quantified using the ``participation ratio'', see \cite{abbott2011interactions} and \methods) of the input distribution that defines the optimal allocation of input power as function of the output noise variance $\sigma^2$. 
  In all panels, colors indicate the degree of non-normality ($\sigma_S$) of the network. Light-colored regions denote 95\% c.i. around the mean, estimated from 50 independent realizations of the $20\times 20$ random matrix $A$, drawn according to \Cref{eq:a_gen_nn}.
 } 
 \vspace{0.1cm}
\label{Fig:rand_nn}
\end{figure*}

To understand the mechanism through which non-normality improves information transmission, we repeat our inspection of optimal power allocation, now for a non-normal network with $\alpha=7$.
In \Cref{Fig:line_nn_rate_full}D, we plot the optimal input covariances (no longer expressed in the eigenbasis of $A$, but in the standard basis of the network's nodes) for various transmission window lengths.
For large transmission windows, including the one that leads to the largest rate $\cR_{\max}$, input power concentrates on the ``source'' nodes (left-most nodes in \Cref{Fig:line_nn_rate_full}A, bottom). This optimal strategy exploits the network's ability to amplify signals as they propagate down the chain towards the ``sink'' (the last node).
Thus, the SNR at the receiver can display large transient increases, whereas its decay could at best be slowed down in normal networks.
For short transmission windows, such a strategy no longer pays off, due to the same tradeoffs as uncovered above for normal networks.
First, the signal transiently builds up into the next transmission epoch, where it no longer is signal but instead contributes noise.
Second, distributing input power unevenly across the $n$ network nodes by favouring the ``source'' nodes reduces the entropy of the input distribution, which fundamentally limits the information rate. 
Together, these drawbacks explain why the source nodes are not particularly favoured over sink nodes when $T$ is small (\Cref{Fig:line_nn_rate_full}D, left), and why, in general, the input power does not concentrate entirely on the first node in the chain, but is generally distributed among the first few.

To further substantiate that non-normality benefits the information capacity, we manipulate the degree of non-normality of the chain network discussed above, this time not by increasing $\alpha$, but through a complementary modification.
Specifically, we morph the non-normal chain discussed above back into a normal network, by chopping the original chain of length $\ell=8$ into sets of shorter chains (\Cref{Fig:line_nn_rate_full}A, top to bottom).
Shorter chains consistently yield smaller information capacity (\Cref{Fig:line_nn_rate_full}B, right), confirming that network non-normality has a positive impact on information transmission. {  We found a similar correlation for the more general class of layered topologies described in our \suppl\ 7. More precisely, for these networks increasing the depth of connected layers has a provably beneficial effect on the communication performance.

\subsection*{How noise shapes the optimal architecture}

The results presented so far show that non-normal architectures can, in principle, outperform normal networks as information transmission media.
These results were obtained for fixed input SNR, and we now show that non-normality is all the more beneficial as the SNR is poor.
To show this, we revisit the chain architecture of the previous section (\Cref{Fig:line_nn_rate_full}A) and systematically vary $\sigma^2$, the amplitude of the noise at the receiver (\Cref{Fig:noise_vs_nn_line}).

In the low-noise regime, non-normality has little impact on information transmission, whether the network is made non-normal by increasing its directionality (\Cref{Fig:noise_vs_nn_line}A) or by increasing the length of its chains (\Cref{Fig:noise_vs_nn_line}B). {  In fact, for any $A$, when $\sigma^2$ is small we have~(cf. \suppl\ 4)
\begin{equation}
\cR_T \approx -\frac{1}{\ln 2}\tr(A),
\label{eq:smallnoise}
\end{equation}
which shows that in the low-noise regime the rate depends on the spectrum of $A$ only.
For the chain network, \Cref{eq:smallnoise} reduces to $\cR_T \approx \frac{\gamma n}{\ln 2}$, which is independent of $\alpha$ and $\ell$.}
For large enough $\sigma^2$, however, increasing $\alpha$ or $\ell$ has pronounced benefits on the maximum information rate $\cR_\text{max}$ (\Cref{Fig:noise_vs_nn_line}A-B).
In contrast, modifications of the parameters of the normal network ($\alpha=1$) that affect the eigenvalues without causing any departure from normality have close to no impact on the information rate.
Specifically, changing the decay rate $\gamma$ of the single nodes is only beneficial in the low-noise regime (\Cref{Fig:noise_vs_nn_line}C), corroborating the conclusions drawn from \Cref{eq:exact_rate_normal} above.
The same equation also predicts that changing the overall coupling strength $\beta$ (while keeping the directionality $\alpha$ constant) have no effect on $\cR_\text{max}$ (not~shown).

From our analysis of this simple architecture, we conclude that network non-normality 
can greatly enhance information transmission in the low SNR regime. In fact, we were able to show that non-normality can (in theory) cancel the effect of noise altogether (\suppl\ 7).  Specifically, it holds
\begin{equation}
 \lim_{\alpha\to \infty}\cR_{T} = -\frac{1}{\ln 2}\tr(A)=-\frac{\gamma n}{\ln 2}.
\label{eq:alphainf}
\end{equation}
\Cref{eq:alphainf} implies that, no matter how poor the SNR is, by increasing the degree of non-normality of the network via the directionality strength $\alpha$ we get arbitrarily close to the maximum information rate achievable in the \emph{noiseless} regime (by any network with identical value of $\text{tr}(A)$; \Cref{Fig:noise_vs_nn_line}A, horizontal dashed red line).

Intriguingly, this result does not only hold for the simple line architecture described above, but also for more complex class of ``layered''
 with  the free parameter $\alpha$  summarising departure from normality in terms of directionality strength between layers (\suppl\ 7).
In this family of models, as in the linear chain, the detrimental effect of output noise (however large) can be annihilated entirely by making the network sufficiently non-normal (by increasing $\alpha$).
In this limit of strong non-normality, the network effectively behaves as a one-dimensional  
channel with decay rate $|\text{tr}(A)|$, and indeed achieves an information rate equal to that of any network with the same $\text{tr}(A)$ in the \emph{absence of output noise} (\Cref{eq:smallnoise}).

Finally, we investigated how the noise level shapes the optimal architecture via an optimization approach. More precisely, we numerically computed the network architecture optimizing the maximum information rate $\mathcal{R}_\text{max}$ with $10$ nodes, bounded network weights and different values of the noise variance $\sigma^2$ (\suppl\ 8).  From our numerical analysis, it turns out that as $\sigma^2$ grows optimal networks become increasingly similar to a purely (hidden or effective) feedforward chain of maximal length, with approximately all of the input power allocated to the first nodes of the chain. This further corroborates our claim that non-normality is crucial for enhancing the communication performance of a network in~the~high-noise~regime.

\subsection*{Generalization to heterogeneous topologies}

Although the formulae we have derived regarding the information capacity of linear networks hold for arbitrary topology, most of the results presented so far were based either on highly simplified, small, and structured architectures (\Cref{Fig:line_nn_rate_full}A), or on networks that deviated from normality in a highly structured way (\suppl\ 7).
To assess the generality of our results, we now study larger and more heterogeneous networks whose departure from normality we can also control.
Specifically, we generate random connectivity matrices $A$ following 
\cite{hennequin2014fast} as:
\begin{equation}
\label{eq:a_gen_nn}
A = (-I + S)P.
\end{equation}
Here $P$ is a random positive definite matrix drawn from the inverse Wishart distribution (\methods), and $S$ is a random skew-symmetric matrix whose (upper-triangular) elements are drawn independently from a normal distribution with zero mean and variance $\sigma_S^2$.
It is easily shown that any state matrix $A$ drawn according to \Cref{eq:a_gen_nn} implies stable network dynamics, despite the network graph showing apparent disorder with connections of arbitrary average magnitude (e.g.\ there is no limit to the norm of $P$ and $S$). The degree of network non-normality is set by the parameter $\sigma_S$: when $\sigma_S=0$, $A$ is symmetric, hence normal; as $\sigma_S$ increases, $A$ departs further from normality {  (cf.~\methods\ and \suppl\ 6)}.
We calculated the maximum rate of such networks for various degrees of non-normality, and found a similar interplay between network non-normality, transmission window, and input SNR as in the simplified architecture of \Cref{Fig:line_nn_rate_full,Fig:noise_vs_nn_line}.
Specifically, non-normality results in greater maximum rates realized by non-zero optimal transmission windows (\Cref{Fig:rand_nn}A).
Moreover, these benefits over normal networks only arise in the low SNR regime (\Cref{Fig:rand_nn}B).
Finally, enhanced transmission performance at low SNR relies on a low-dimensional allocation of input power (\Cref{Fig:rand_nn}C).

The role of non-normality in information transmission is further illuminated by considering the limit of poor SNR ($\sigma^2 \to \infty$): for any transmission window length $T>0$, the rate decays with growing $\sigma^2$ as (cf. \suppl\ 4)
\begin{align}
  \cR_T \approx \frac{1}{2 \ln 2\, T} \frac{\|B^\top \cO B\|}{\sigma^2},
  \label{eq:lim_SNR_zero}
\end{align}
where $\| B^\top \cO B \|$ represents the maximum total energy that the network can autonomously generate over a time window $T$, for an appropriate encoding of the input packet $u_0$. 
While the momentary magnitude of activity in normal networks can only decay in time (leading to sublinear growth of $\|B^\top \cO B \|$ with $T$, i.e.\ decreasing $\cR_T$ in \Cref{eq:lim_SNR_zero}), non-normal networks have the capacity to transiently amplify certain input codes before the eventual decay of signals implied by collective stability.
This leads to superlinear growth of $\|B^\top \cO B \|$ with $T$, which in turn results in transiently increasing $\cR_T$ peaking at some finite value of $T$ (\Cref{eq:lim_SNR_zero}).

Finally, in deriving \Cref{eq:lim_SNR_zero}, we could also prove that in the limit of large noise $\sigma^2$, the rate $\cR_T$ is realized by effectively one-dimensional inputs, whose distribution lies entirely along the most sensitive input direction (i.e.\ along the initial condition that evokes the largest energy in the window $T$; \suppl\ 4).
In other words, the best way for the network to counteract a large amount of noise is to map every input packet onto a single, maximally amplified input pattern, thus effectively giving up on most of its degrees of freedom. 
This corroborates and strengthens the generality of our findings of \Cref{Fig:line_nn_rate_full}D and \Cref{Fig:rand_nn}C regarding the effective dimensionality of the input distribution in the high-noise~regime.

\section*{Discussion}

In this paper we have proposed a novel framework to model information propagation through networks with arbitrary topology and nodes governed by linear dynamics.
{  These dynamics imply a form of memory in single nodes, giving rise to interference between the activity transient initiated by the presentation of a given input packet, and the activity left over from previous transmissions.
We have  used the notion of Shannon's mutual information to quantify communication performance, and study how the latter depends on the network architecture}.
Our analysis has shown that the qualitative effects of graph connectivity on communication are largely determined by a property that is often overlooked: the degree of non-normality of the network's (weighted) adjacency matrix.
In particular, we have shown that normal networks perform poorly in the presence of large readout noise at the receiver.
In contrast, non-normal networks exhibit more favorable communication properties, including the ability to entirely cancel out the effect of readout noise provided the input packets are appropriately encoded, and the adjacency matrix is sufficiently non-normal.
Interestingly, non-normal networks appear ubiquitous, with strong non-normality having been found in foodwebs, transport, biological, social, communication, and citation networks~\cite{ALC18}. In addition, we mention that, besides information transfer, non-normality turns out to be the key to explaining and understanding a variety of other equally important phenomena. For instance, the process of pattern formation in natural and biological systems \cite{Asllani18,Muolo}, the selective amplification of cortical activity patterns in the brain \cite{Murphy09}, and the emergence of giant oscillations in noise-driven dynamical~systems~\cite{Fanelli,NicolettiA,NicolettiB}.

\smallskip

To further highlight the impact and potential practical relevance of our findings, we have used our framework to analyze the communication performance of the neuronal network of the nematode \emph{Caenorhabditis elegans}. We focused on the (weighted and directed) chemical synapse network described in  \cite{Varshney,WormAtlas}, and examined the linearized and stabilized network dynamics of the neuronal membrane potentials (\methods).  The network, which is illustrated in \Cref{Fig:c_elegans}A, comprises 279 neurons (divided into 88 sensory neurons, 82 interneurons, and 107 motor neurons) recurrently coupled through 2194 inhibitory/excitatory synaptic connections. We first wondered whether the non-normality of this directed biological connectome had the beneficial impact on communication that we have documented here for artificial networks.  We thus compared its information rate $\mathcal{R}_T$ (as a function of the transmission window $T$) with that of a symmetrized version (implying normal $A$), as well as a randomized ensemble wherein the direction of each existing coupling in the connectome is reversed with probability $1/2$.  Both manipulations induce a significant drop in $\mathcal{R}_T$ from the real network (\Cref{Fig:c_elegans}B), indicating that the \emph{C.~elegans} connectome is non-normal in a way that benefits information transmission as shown in this paper.  We next wondered if the network's non-normal structure is likely to be exploited for communication by these organisms. We reasoned that communication would naturally flow from sensory neurons to motor neurons, and that the network should therefore display good communication (in our framework) if, and only if, the input matrix $B$ were to select sensory neurons while the output matrix $C$ were to read out motor neurons. We found that this is indeed the case (compare \Cref{Fig:c_elegans}C green and blue). Strikingly, also, the symmetrized version of the connectome is almost unable to communicate information from sensory to motor nodes (Figure \Cref{Fig:c_elegans}B, red). Although preliminary, these numerical findings could shed light on the actual functioning of the \emph{C.~elegans} neuronal circuit and behavioural responsiveness to external stimuli. More generally, we expect that our theoretical framework could be used to understand and explain the emergence of certain topological structures in biological networks, and to identify their intrinsic communication pathways.

\begin{figure*}[t]
\begin{center}
\includegraphics[scale=0.465]{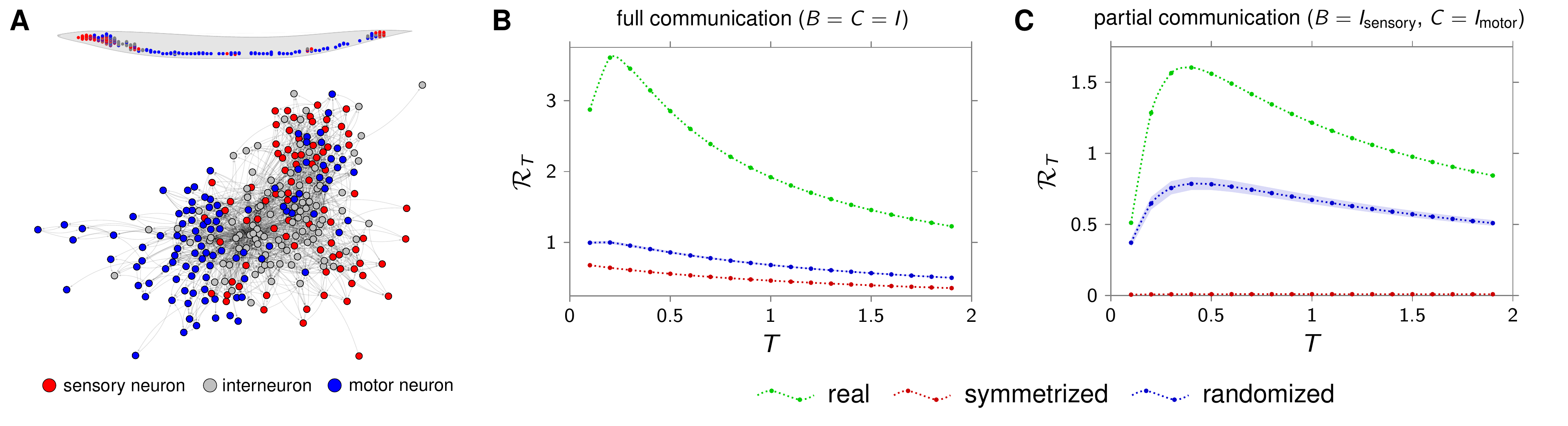} 
\end{center}
\caption{
{\bfseries Information rate of \emph{C.~elegans} network.}
{\bfseries (A)}~Schematic of the chemical synapse network of the \emph{Caenorhabditis elegans} nematode. The connectivity data and soma position of the neurons are taken from \cite{Varshney,WormAtlas}. 
   {\bfseries (B)}~Plot of the information rate $\mathcal{R}_{T}$ versus transmission time $T$ in the full communication setting (matrices $B=C=I$) for the chemical synapse network $A$ of the \emph{C.~elegans} (green curve), its symmetrized version $(A+A^{\top})/2$ (red curve), and a randomized version in which each pair of entries $(A_{ij},A_{ji})$ has been swapped with probability $1/2$ (blue curve). 
  \textbf{(C)}~Plot of the information rate $\mathcal{R}_{T}$ versus transmission time $T$ in a partial communication setting (matrices $B\ne I, C\ne I$). The green curve is the rate of the real network $A$, with $B$ and $C$ selecting the 88 sensory and 107 motor neurons, respectively. The red curve is the rate of the symmetrized network described by matrix $(A+A^{\top})/2$, with matrices $B$ and $C$ as above. The blue curve is the rate of the real network described by matrix $A$, with matrices $B$ and $C$ randomly selecting 88 input nodes and 107 output nodes, respectively. The sets of input and output nodes are chosen to be non-overlapping. In all plots, the networks have been stabilized, by shifting their spectrum by a scalar matrix $\gamma I$, $\gamma\in\mathbb{R}$, so that the real part of the largest eigenvalue is $-0.1$, and the variance of the output noise is set to $\sigma^{2}=1$. In the randomized scenarios, the dash-dotted curves represent the mean over 100 realizations and the light-blue regions denote~95\%~c.i.~around~the~mean.
 } 
 \vspace{0.1cm}
\label{Fig:c_elegans}
\end{figure*}

\smallskip

In the paper we focused on weighted networks, and regarded the weights (and, precisely, their directionality and magnitude) as the main factors influencing network non-normality. However, non-normal architectures can also emerge in unweighted networks, e.g., in networks with heterogeneous outdegree/indegree distributions. It would therefore be interesting to investigate what the most relevant features impacting non-normality in unweighted networks are, and to what extent these features affect communication performance. Further, in our framework, noise is modelled as the combined effect of readout noise and an internal, structured source of noise arising from inter-symbol interference. Investigating how different noise models could affect our analysis and results represents a compelling direction of future research. Also, noise could play an active role in the information transfer process as the input source of the communication channel. This change of perspective could lead to an information-theoretic interpretation of the findings of \cite{Fanelli,NicolettiA,NicolettiB}, wherein non-normality has been linked with the emergence of amplified oscillations in noise-driven interconnected  non-linear systems.

As is well known in the theory of non-normal matrices and operators \cite{Trefethen2005}, strong departure from normality often implies heightened sensitivity to structural perturbations | for example, the random addition/deletion of nodes or edges in a graph.
This suggest a generic trade-off between communication performance and resilience, which would be interesting to study further.
For example, we note that in the low-noise regime where normal networks can perform just as well as non-normal ones, constraints on robustness would favour normal networks. A similar trade-off has been identified recently in \cite{PFZZ18} where network resilience was shown to be generically at odds with network controllability.

Our work may also offer new perspectives on memory and information storage.
Information transmission and storage are very similar problems: communication is transmission through space, while memory is transmission through time.
Indeed, these two problems admit very similar models, are often both approached using the tools of information theory \cite{Ganguli08,Toyoizumi12,Ganguli10}, and may interact in the context of network in ways that would be interesting to investigate further.
Preliminary intuitions suggest that they may benefit each other:
in our communication model, for example, inter-symbol interference could be {  reduced} if one could keep a memory of decoded past packets, and subtract their individual contributions to the momentary network activity at any time.
Conversely, communication may improve memory.
An obvious example is the oral tradition in human communities, where transmission of information from generation to generation emerges as a way to overcome the finite memory- (and indeed, life-) span~of~individuals.

\small

\section*{\label{sec:methods}Materials and Methods}
\vspace{-0.25cm}

\paragraph{Gramian matrices and numerical computation of information capacity and rate.}
The observability Gramian over the interval $[0,T]$ of the system in \Cref{eq:linearsys1} is defined as
\begin{align}\label{eq:Ogram}
\cO=\int_0^T e^{A^\top t}C^\top C e^{At}\, \text{d} t,
\end{align}
and can be numerically evaluated via numerical integration of the matrix-valued differential equation:
\begin{align}\label{eq:Ogram-num}
\dot{X}(t)=A^\top X(t) + X(t) A + C^\top C,
 \end{align}
subject to the initial condition $X(0)=0$ \cite{H09}.
The infinite-horizon controllability Gramian of the dynamics in \Cref{eq:linearsys1} discretized with sampling time $T$ as
 \begin{align}\label{eq:Wgram}
 \cW=\sum_{k=0}^{\infty} e^{A kT} B\Sigma B^\top e^{A^\top kT}.
 \end{align}
 and can be computed as the solution of the discrete-time algebraic Lyapunov equation: 
 \begin{align}\label{eq:Wgram-num}
 X - e^{AT} X e^{A^\top T} = B\Sigma B^\top.
 \end{align} 
 In the numerical evaluation of the capacity and rate, the Gramians \eqref{eq:Ogram} and \eqref{eq:Wgram} has been computed via \eqref{eq:Ogram-num} and \eqref{eq:Wgram-num}, respectively. 
For vector-valued inputs ($m\ge 2$), the solution of the optimization problem in \Cref{eq:cap,eq:rate} has been numerically carried out in Python using optimization routines from the Pymanopt library \cite{Pymanopt}, together with automatic differentiation techniques provided by Autograd \cite{Autograd}. If $B=C=I$ and $A$ is normal, the solution is unique and admits a closed-form expression in terms of eigenvalues of $A$ (\suppl\ 5). More generally, if $C^{\top}C\succcurlyeq e^{A^{\top}T}C^{\top}Ce^{AT}$, then the optimization in \Cref{eq:cap,eq:rate} is convex (\suppl\ 3), and so convergence to the maximum is always guaranteed using trust-region or steepest descent methods. Otherwise, the problem turns out to be, in general, non-convex, and, in order to avoid local maxima, we ran the latter routines several times ($10^2$-$10^3$), starting from different random initializations, and selected the largest outcome. 

\vspace{-0.25cm}
\paragraph{Generation of random non-normal matrices and participation ratio.}
In \Cref{eq:a_gen_nn}, the skew-symmetric matrix $S\in\R^{n\times n}$ has been generated as $S=L-L^\top$, with $L_{ij}\sim \cN(0,\sigma_S^2)$ for $i<j$, and $L_{ij}=0$ otherwise. The positive definite matrix $P\in\R^{n\times n}$ has been drawn from the inverse Wishart distribution with scale matrix $\omega^{-2}I$ and $\nu$ degrees of freedom. We chose $\omega^{-2}=\nu-n-1$,  $\nu=24+n$, in order to guarantee sufficient heterogeneity in the eigenvalues of $P$ \cite{hennequin2014fast}. With this choice, it can be shown that $\sigma_S^2$ correlates well with standard measures of matrix non-normality { (\suppl\ 6)}. Following \cite{abbott2011interactions}, given a positive definite matrix $A\in\mathbb{R}^{n\times n}$ with eigenvalues $\{\lambda_i\}_{i=1}^n$, we define the participation ratio
\begin{align}
    n_{\text{eff}} = \frac{\left(\sum_{i=1}^n \lambda_i \right)^2}{\sum_{i=1}^n \lambda_i^2}.
\end{align}
When applied to the covariance matrix, the participation ratio provides a measure of the effective dimensionality of the underlying random vector.  

\vspace{-0.25cm}

\paragraph{\emph{Caenorhabditis elegans} dataset and network dynamics.} The \emph{C. elegans} connectivity data of \cite{Varshney,WormAtlas} comprise two datasets: the gap junction and chemical synapse wiring diagrams. Since the gap junction dataset does not include link directionality, in our study we focus on the chemical synapse network which possesses clear directionality extracted from electron micrographs. This network consists of 279 neurons. These neurons are categorized in 88 sensory neurons (neurons known to respond to specific environmental conditions), 107 motor neurons (neurons characterized by the presence of neuromuscular junctions), and 82 interneurons (the remainder). The network comprises 2194 synaptic connections. As in \cite{Varshney}, we make the common assumption that GABAergic neurons (26 neurons) make inhibitory synapses, whereas the rest of the neurons form excitatory synapses.  We describe the autonomous dynamics of the chemical synapse network by the following linear system 
\begin{align}\label{eq:celegans}
    \tau \dot x(t) = (A- \gamma I) x(t),
\end{align}
where $x$ is the vector containing the membrane potentials of all neurons around an equilibrium, $A$ is the adjacency matrix of the chemical synapse network,  $\tau = C/g$, and $\gamma=g^{m}/g$. Here, the parameters $C$, $g$, and $g^{m}$ represent the (average) neuronal membrane capacitance, synaptic conductance, and membrane conductance, respectively (see \cite{ferree,Varshney} for further details). In our numerical study, we set $\tau=0.5$ and tune $\gamma$ in order to stabilize the network dynamics \eqref{eq:celegans}. Specifically, we set the largest real part of the eigenvalues to $-0.1$. This yields $\gamma \approx 20$, a value within the physiological range of $g^{m}$ and $g$ \cite{Nicolettielegans}.  However, profiles of $\mathcal{R}_{T}$ qualitatively similar to those in \Cref{Fig:c_elegans}A, B have been obtained for a wide range of values of parameters $\gamma<0$, $\tau>0$, and noise variance $\smash{\sigma^{2}}$.

\section*{\label{sec:methods}Supplementary Materials}

\vspace{-0.25cm}

Note S1: Controllability and observability Gramians.\newline
Note S2: Derivation of the information capacity formula.\newline
Note S3: Properties of the information capacity and rate.\newline
Note S4: Information rate in the low and high noise regime.\newline
Note S5: Information rate of normal networks.\newline
Note S6: Measures of matrix non-normality and network indicators of non-normality.\newline
Note S7: Information rate of a class of non-normal networks.\newline
Note S8: Optimal communication architectures.\newline
Fig.~S1: Non-normality metrics and parameters of chain~network.\newline
Fig.~S2: Non-normality metrics and parameters of heterogeneous network.\newline
Fig.~S3: Construction of ``layered'' non-normal networks.\newline
Fig.~S4: Optimal networks and input covariances as a function of the transmission window $T$. \newline
Fig.~S5: Optimal networks and input covariances as a function of the noise covariance $\sigma^{2}$. \newline
References: \cite{PZB14,BBZ17,liu2016control,DC01,B13,fang1994inequalities,HJ12,NW06}.

\vspace{-0.25cm}
\small 
\bibliography{ref}
\bibliographystyle{ScienceAdvances}

\vspace{-0.5cm}
\small 
\paragraph{Funding.} This work was supported by a Wellcome Trust Seed Award 202111/Z/16/Z (G.H.), the Howard Hughes Medical Institute through a Janelia Graduate Research Fellowship~(V.R.), and partially by MIUR (Italian Minister for Education)  under the initiative ``Departments of Excellence''~(Law 232/2016)~(G.B., S.Z.).
 
 \vspace{-0.5cm}

\paragraph{Author Contributions.} G.B., V.R., G.H, and S.Z.\ all contributed to the conceptual and theoretical aspects of the study, and wrote the manuscript.
 G.B.\ carried out the numerical simulations and made the figures.
 G.B.\ and S.Z.\ wrote the supplementary material. 

\vspace{-0.5cm}

\paragraph{Competing interests.} The authors declare no competing~interests.

\vspace{-0.5cm}

\paragraph{Data and materials availability.} The code used in this study is freely available in the public GitHub repository: \href{http://github.com/baggiogi/commun_complex_networks}{http://github.com/baggiogi/commun\_complex\_networks}. The \emph{C. elegans} connectivity data is freely available at \href{http://www.wormatlas.org/neuronalwiring.html}{http://www.wormatlas.org/neuronalwiring.html}.

\end{document}



\maketitle

\blfootnote{\textsuperscript{$\ast$}Corresponding author: \href{mailto:sandro.zampieri@unipd.it}{sandro.zampieri@unipd.it}.} \blfootnote{\textsuperscript{$\dagger$}Equal contributions.}

\tableofcontents

\newpage 


\section{Controllability and observability Gramians}

Consider the continuous-time linear time-invariant system
\begin{align}\label{eq:linsys}
\begin{split}
    \frac{\de x(t)}{\de t}&=Ax(t)+Bu(t), \\
    y(t)&=Cx(t),
\end{split}
\end{align} 
where $t\ge 0$ and $x(0)=x_0\in\R^n$. Here, $x(t)\in\R^n$, $u(t)\in\R^m$, and $y(t)\in\R^p$ denote the state, input, and output of the system at time $t$, respectively, and $A$, $B$, and $C$ are matrices of suitable dimensions.
The controllability Gramian of the system over the interval $[0,T]$, $T>0$, is defined as \begin{align}\cW=\int_0^T e^{A t}BB^\top e^{A^\top t} \, \text{d} t,\end{align}  whereas the observability Gramian of the system over the interval $[0,T]$, $T>0$, as \begin{align}\cO=\int_0^T e^{A^\top t}C^\top C e^{At}\, \text{d} t.\end{align}
The above-defined Gramians matrices $\cW$ and $\cO$ are always positive semidefinite, and they are related to the controllability and observability properties of the linear system in \Cref{eq:linsys}. In particular, the system is controllable if and only if $\cW$ is positive definite, and observable if and only if $\cO$ is positive definite (see e.g. \cite{H09} for further details). 
Furthermore, if $A$ is Hurwitz stable (that is, all the eigenvalues of $A$ have negative real part), then $\cW$ and $\cO$ are well-defined for $T\to\infty$, and they correspond to the (unique) positive semidefinite solution of the continuous-time Lyapunov equations $AX+XA^\top = -BB^\top$ (controllability Gramian) and $A^\top X+XA = -C^\top C$ (observability Gramian). From an intuitive viewpoint, $\cW$ is related to the energy needed to steer the system from an initial state $x_0$ to a desired target state $x_f$ at time $T$, i.e. $x(T)=x_f$, whereas $\cO$ to the energy of the system's free response evoked by a certain initial state $x_0$ in the interval $[0,T]$, see e.g. \cite{PZB14,BBZ17,liu2016control}. 

Similar definitions/interpretations hold for systems governed by discrete-time dynamics
\begin{align}
\begin{split}
    x(k+1)&=Ax(k)+Bu(k),\\
    y(k)&=Cx(k),
\end{split}
\end{align} 
where $k\in \N_{>0}$ and $x(0)=x_0\in\R^n$.
In this case, the controllability and observability Gramians over the interval $[0,T]$, $T\in\mathbb{N}$, are given, respectively, by \begin{align}\cW=\sum_{k=0}^{T-1} A^k BB^\top (A^\top)^k\ \ \  \text{ and }\ \ \ \cO=\sum_{k=0}^{T-1} (A^\top)^k C^\top C A^k.\end{align}
If $A$ is Schur stable (all the eigenvalues of $A$ have modulus strictly less than one), then the infinite-horizon ($T\to \infty$) controllability and observability Gramians coincide with the positive semidefinite solution of the discrete-time Lyapunov equation $AXA^\top-X = -BB^\top$ and $A^\top XA - X = -C^\top C$, respectively.

\section{Derivation of the information capacity formula}

Here and throughout the paper, we indicate with $\det(X)$ and $\tr(X)$ the determinant and trace of matrix $X$, respectively, and $X\succcurlyeq 0$ ($X\succ 0$) means that $X$ is positive semidefinite (positive definite, respectively). Further, given two continuous random variables $x$ and $y$, we let $h(x)$ and $h(x|y)$ denote, respectively, the differential entropy of $x$ and the conditional differential entropy of $x$ given $y$. The mutual information between two continuous random variable $x$ and $y$ is given by $\mathcal{I}(x;y) = h(x) - h(x|y)$ \cite{CT12}.

We remark that the mutual information between the input and output messages of a communication channel, further optimized over all power-limited input distributions, coincides with the maximum achievable rate of reliable information transmission (Shannon's channel capacity) when the channel is \emph{memoryless}, i.e., when past inputs do not affect the currently transmitted message \cite{CT12}. This is not, however, the case in our scenario because of inter-symbol interference. Moreover, even in the memoryless case, to achieve Shannon's capacity the encoder and the decoder must keep memory of previous transmissions. In most applications of network transmission (as those mentioned in the main text), instead, memory is not available, and so reconstruction has to be performed on a~packet-by-packet~basis.

To derive the capacity formula in Equation 5 of the main text, we first address the case $B=I$, and then we show how to extend the argument to the general case of rectangular $B$'s.
With reference to the communication channel described in the main text, we focus, without any loss of generality, on the transmission delivered at time $t=0$.  
The total information signal in the time window $[0,T]$ will be the superposition of the signals $y_{k}(t)$'s in the same window, namely $\sum_{k\in\mathbb{Z}}y_{k}(t)$, $0\le t\le T$. In the latter summation, the signal that contains the ``useful'' information is given by $y_{0}(t)$, $0\le t\le T$. Let $\mathcal{L}_{2}^{p}[0,T]$ be the Hilbert space of square integrable functions over the interval $[0,T]$ equipped with the inner product $\langle f,g\rangle_{\mathcal{L}_2}=\int_{0}^{T}f(t)g(t)\,\mathrm{d} t$, and note that $y_{0}(t)$, $0\le t\le T$, belongs to the finite-dimensional subspace $\mathcal{Q}$ of $\mathcal{L}_{2}^{p}[0,T]$ generated by the functions $\{Ce^{A t}e_{i}, \ t\in[0,T]\}_{i=1}^{n}$, where $\{e_i\}_{i=1}^n$ denotes the canonical basis vectors of $\R^n$. Thus, $y_{0}(t)$ can be written as
\begin{align}
	y_{0}(t)=\sum_{i=1}^{M}y_{i}f_{i}(t),
\end{align}
where $\{f_{i}(t)\}_{i=1}^{M}$ is any orthonormal basis in $\mathcal{Q}$ 
 and
\begin{align}
	y_{i}&:=\langle f_{i}(t),y_{0}(t)\rangle_{\mathcal{L}_2}=\int_{0}^{T}f_{i}^{\top}(t)y_{0}(t)\,\mathrm{d} t=\underbrace{\int_{0}^{T}f_{i}^{\top}(t)Ce^{At}\,\mathrm{d}  t}_{\displaystyle =:F_{i}^{\top}}u_{0}.
\end{align}
Let us define $F:=[F_{1},\dots,F_{M}]^{\top}$ and $Y_{0}:=[y_{1},\dots,y_{M}]^{\top}$. For all $u_{0}\in\R^{n}$, it holds
\begin{align}
	\langle y_{0}(t),y_{0}(t)\rangle_{\mathcal{L}_2} &= u_{0}^{\top} \int_{0}^{T} e^{A^{\top}t}C^{\top} Ce^{At}\,\mathrm{d} t\,  u_{0} =Y_{0}^{\top} Y_{0} = u_{0}^{\top} F^{\top} F  u_{0}.
\end{align}
The latter equation implies that $F^{\top}F=\cO$ where $\cO=\int_{0}^{T}e^{A ^{\top}t}C^{\top}Ce^{A t}\,\de t$ is the observability Gramian of the pair $(A,C)$ over the interval $[0,T]$. 
The covariance between two components $y_{h}$ and $y_{\ell}$, $h,\ell=1,2,\dots,M$, is given by
$
\E[y_{h}y_{\ell}] = \E\left[F^{\top}_{h}u_{0}u_{0}^{\top}F_{\ell}\right]=F^{\top}_{h}\Sigma F_{\ell},
$
where $\Sigma:=\E[u_0 u_0^\top]$.
From this fact, it follows that the covariance of the useful signal $y_{0}(t)$ is
\begin{align} \label{eq:sigcov}
	\Sigma _{y_{0}}:=\E\left[Y_{0}Y_{0}^{\top}\right]=F\Sigma F^{\top}.
\end{align}
The overall channel noise, here denoted by $r(t)$, is modelled as the sum of two contributions, namely: 
\begin{enumerate}[(i)]
\item the additive Gaussian white noise $n(t)$, and 
\item the interference term $i(t)$ due to inter-symbol interference.  
\end{enumerate}
The noise term $r(t)=n(t)+i(t)$ can be written as, w.r.t.~the previously introduced orthonormal basis $\{f_{i}(t)\}_{i=1}^{M}$ of $\mathcal{Q}$,
\begin{align}
	r(t)=\sum_{i=1}^{M}r_{i}f_{i}(t) + r_{\perp}(t),
\end{align}
where $r_{\perp}(t)$ belongs to the orthogonal complement of $\mathcal{Q}$ and 
\begin{align}
	r_{i}&:=\langle f_{i}(t),r(t)\rangle_{{\cal L}_{2}} = \int_{0}^{T}f_{i}^{\top}(t) \left(Ce^{At}\sum_{k=1}^{\infty} e^{AkT}u_{-k} +n(t)\right) \de t = F_{i}^{\top}\sum_{k=1}^{\infty} e^{AkT}u_{-k} + n_{i},
\end{align}
with $n_{i}:=\int_{0}^{T}f_{i}(t)^{\top} n(t)\, \de t$. The covariance between two noise components $r_{h}$, $r_{\ell}$, $h,\ell=1,2,\dots,M$, is given by
\begin{align}
\E[r_{h}r_{\ell}] &= F_{h}^{\top} \sum_{k=1}^{\infty} e^{AkT}\E[u_{-k}u^{\top}_{-k}]e^{A^{\top}kT} F_{\ell}  +\sigma^{2}\int_{0}^{T}\int_{0}^{T}f_{h}^{\top}(t) f_{\ell}(\tau) \delta(t-\tau)\, \de t\, \de \tau =F_{h}^{\top} \sum_{k=1}^{\infty} e^{AkT} \Sigma e^{A^{\top}kT} F_{\ell} + \sigma^{2}\delta_{h,\ell},
\end{align}
where $\delta_{h,\ell}$ denotes the Kronecker delta function and we implicitly used the fact that $\{u_k\}$ are i.i.d. random variables, and $u_{k}$ and $n_i$ are independent for all $k$ and $i$.
Therefore, by defining $R:=[r_{1},\dots,r_{M}]^{\top}$, we have that the covariance of the noise component in $\mathcal{Q}$ is given by
\begin{align}\label{eq:noisecov}
	\Sigma _{r}:=\E\left[RR^{\top}\right] &= F \sum_{k=1}^{\infty} e^{AkT}\Sigma e^{A^{\top}kT} F^{\top} + \sigma^{2}I = F(\cW -\Sigma )F^{\top} + \sigma^{2}I,
\end{align}
where $\cW:=\sum_{k=0}^{\infty}e^{A kT}\Sigma  e^{A ^{\top}kT}$ is the infinite-horizon controllability Gramian of the discretized pair $(e^{AT},\Sigma^{1/2})$. 

By exploiting the above finite-dimensional representation of the output signal and noise term and the covariance expressions in \Cref{eq:sigcov,eq:noisecov}, we can now compute the mutual information (as measured in bits) between the channel input $u_0$ and the corrupted output trajectory over the window $[0,T]$, $\tilde Y_0$,
\begin{align}
	\mathcal{I}_T(u_0, \tilde Y_0) &=  h(\tilde Y_0) - h(\tilde Y_0|u_0) =  h(Y_0+R) - h(Y_0+R|u_0) = h(Y_0+R) - h(R) \notag\\
	& = \frac{1}{2} \log_{2} \frac{\det(\Sigma_{y_{0}}+\Sigma_{r})}{\det\Sigma_{r}} =\frac{1}{2}\log_{2} \frac{\det(\sigma^{2}I+\cO\cW)}{\det(\sigma^{2}I+\cO(\cW -\Sigma ))},
\end{align}
where we used the properties of the differential entropy, the fact that $Y_0$ and $R$ are independent Gaussian random variables \cite[Ch.~8]{CT12}, and, in the last step, the similarity invariance of the determinant and the fact that $F^{\top} F=\cO$. 
Finally, the previous expression yields the expression of the information capacity with power constraint $P$,
\begin{align}
	\cC_T&=\frac{1}{2}\, \max_{\Sigma\succcurlyeq 0, \tr\Sigma  \le P} \mathcal{I}_T(u_0, \tilde Y_0) =\frac{1}{2}\, \max_{\Sigma\succcurlyeq 0, \tr\Sigma  \le P}\log_{2} \frac{\det(\sigma^{2}I+\cO\cW)}{\det(\sigma^{2}I+\cO(\cW -\Sigma ))}.
\end{align}
This concludes the proof for the case $B=I$.

For general input matrices $B\in\R^{n\times m}$, the search space $\{\Sigma\in\R^{n\times n},\, \Sigma \succcurlyeq 0,\, \tr\,\Sigma\le P\}$ in the maximization of the previously derived formula should be replaced by
$
	\{B\Sigma B^{\top}\,:\, \Sigma\in\R^{m\times m}, \ \Sigma\succcurlyeq 0, \ \tr\,\Sigma\le P\}.
$
Hence, it holds
\begin{align}\label{eq:cap}
	\cC_T=\frac{1}{2}\,\max_{\substack{\Sigma\succcurlyeq 0, \tr\,\Sigma  \le P}} \log_{2} \frac{\det(\sigma^{2}I+\cO \cW)}{\det(\sigma^{2}I+\cO(\cW -B\Sigma B^{\top} ))},
\end{align}
where $\cW:=\sum_{k=0}^{\infty}e^{A kT}B\Sigma B^{\top} e^{A ^{\top}kT}$ is the infinite-horizon controllability Gramian of the discretized pair $(e^{AT},B\Sigma^{1/2})$.

To conclude, we show that the constraint $\tr\,\Sigma  \le P$ in \Cref{eq:cap} can be replaced by $\tr\,\Sigma = P$. To this end, Let us define
\begin{align}
	f_{T}(\Sigma):=\log_{2}\frac{\det(\sigma^{2}I+\cO\cW)}{\det(\sigma^{2}I+\cO(\cW-B\Sigma B^{\top}))}.
\end{align}
We will show that $f_{T}(\alpha\Sigma)$ is a monotonically increasing function of $\alpha\ge 0$, that is, for any $\alpha_{1},\, \alpha_{2}\ge 0$ such that $\alpha_{1}\le \alpha_{2}$, it holds $f_{T}(\alpha_{1}\Sigma)\le f_{T}(\alpha_{2}\Sigma)$. The latter fact clearly implies that the optimal $\Sigma$ maximizing $f_{T}(\Sigma)$ under the constraint $\tr\,\Sigma\le P$, must satisfy the latter constraint with equality.  We have
\begin{align}
	f_{T}(\alpha \Sigma) &= \log_{2}\frac{\det(\sigma^{2}I+\alpha\cO\cW)}{\det(\sigma^{2}I\alpha\cO(\cW-B\Sigma B^{\top}))} \notag\\
	&= \log_{2}\frac{\det(\alpha\cO^{1/2}\cW\cO^{1/2} +\sigma^{2}I)}{\det(\alpha\cO^{1/2}(\cW-B\Sigma B^{\top})\cO^{1/2} +\sigma^{2}I)}\notag \\
	&=\sum_{i=1}^{N}\log_{2}(\alpha\lambda_{i}+\sigma^{2})-\sum_{i=1}^{N}\log_{2}(\alpha\mu_{i}+\sigma_{r}^{2}),
\end{align}
where $\{\lambda_{i}\}_{i=1}^{n}$ and $\{\mu_{i}\}_{i=1}^{n}$ denotes the ordered eigenvalues of $\cO^{1/2}\cW\cO^{1/2}$ and $\cO^{1/2}(\cW-B\Sigma B^{\top})\cO^{1/2}$, respectively. Taking the derivative w.r.t. $\alpha$ of the previous expression, we get
\begin{align}
	\frac{\de f_{T}(\alpha \Sigma)}{\de\alpha}=\frac{1}{\ln2}\sum_{i=1}^{N}\frac{\lambda_{i}}{\alpha\lambda_{i}+\sigma^{2}}-\frac{1}{\ln2}\sum_{i=1}^{N}\frac{\mu_{i}}{\alpha\mu_{i}+\sigma^{2}}.
\end{align}
Since $\cO^{1/2}\cW\cO^{1/2}\succcurlyeq \cO^{1/2}(\cW-B\Sigma B^{\top})\cO^{1/2}$, then $\lambda_{i}\ge \mu_{i}$, $i=1,2,\dots,n$. This yields
\begin{align}
	\frac{\de f_{T}(\alpha \Sigma)}{\de \alpha}=\frac{1}{\ln2}\sum_{i=1}^{N}\frac{\lambda_{i}-\mu_{i}}{(\lambda_{i}\alpha+\sigma^{2})(\mu_{i}\alpha+\sigma^{2})}\ge 0,
\end{align}
which, in turn, implies that $f_{T}(\alpha\Sigma)$ is a monotonically increasing function of $\alpha\ge 0$, as required.

\section{Properties of the information capacity and rate}

In this Supplementary Note, we collect and discuss some key properties of the information capacity $\cC_T$ and information rate $\cR_T$. The first result asserts that $\cC_T$ (and, therefore, $\cR_T$) does not independently depend on the available power $P$ and readout noise variance $\sigma^2$, but on the ``signal-to-noise'' ratio $P/\sigma^2$ only.

\begin{prop}[Scaling invariance of $\cC_{T}$ w.r.t.~$P$ and $\sigma^{2}$] \label{prop:SNR} For all $\alpha>0$, it holds
\begin{align}
\cC_{T}(P,\sigma^{2}) = \cC_{T}(\alpha P,\alpha\sigma^{2}),
\end{align}
where we made explicit the dependence of $\cC_{T}$ on $P$ and $\sigma^{2}$.
\end{prop}
\begin{proof} 
First observe that if we replace $\Sigma$ and $\sigma^{2}$ with $\alpha \Sigma$ and $\alpha \sigma^{2}$, respectively, then $\cW$ is replaced with $\alpha \cW$. This in turn implies that the value of
\begin{align}
f_{T}(\Sigma,\sigma^{2}):=\log_{2}\frac{\det(\sigma^{2}I+\cO\cW)}{\det(\sigma^{2}I+\cO(\cW-B\Sigma B^{\top}))}
\end{align}
is not affected by this change of variables. If $\Sigma^{\star}$ is the input covariance maximizing $f_{T}(\Sigma,\sigma^{2})$ under the constraint $\tr\,\Sigma= P$, it holds
\begin{align}
\cC_{T}(\alpha P,\alpha\sigma^{2}) \ge f_{T}(\alpha\Sigma^{\star},\alpha\sigma^{2})=f_{T}(\Sigma^{\star},\sigma^{2})= \cC_{T}(P,\sigma^{2}).
\end{align}
On the other hand, let $\Sigma^{\star}\succcurlyeq 0$ now denote the input covariance maximizing $f_{T}(\Sigma,\alpha \sigma^{2})$ under the constraint $\tr\,\Sigma= \alpha P$. It holds
\begin{align}
\cC_{T}(P,\sigma^{2}) \ge f_{T}\left(\frac{1}{\alpha}\Sigma^{\star},\sigma^{2}\right)=f_{T}(\Sigma^{\star},\alpha\sigma^{2})= \cC_{T}(\alpha P,\alpha \sigma^{2}).
\end{align}
Therefore, it must be $\cC_{T}(\alpha P,\alpha\sigma^{2})=\cC_{T}(P,\sigma^{2})$.
\end{proof}

We now investigate the convexity properties of the maximizing function in \Cref{eq:cap}, namely
\begin{align}\label{eq:max-func}
 f_{T,\sigma^2}(\Sigma) := \frac{1}{2} \log_2 \frac{\det \left(\sigma^2I+ \cO \cW \right)}{\det \left(\sigma^2I+  \cO (\cW-B\Sigma B^{\top}) \right)},
\end{align}
where we used the subscripts $T$ and $\sigma^2$ to make explicit the dependence of $f$ on the transmission time window and output noise variance, respectively.
 The following result provides a sufficient condition under which the $f_{T,\sigma^2}(\Sigma)$ is a concave function of $\Sigma\succcurlyeq 0$.
\begin{prop}[Concavity of $f_{T,\sigma^2}(\Sigma)$] \label{prop:concave} Assume that the pair $(A,C)$ is observable. If the following condition is satisfied
\begin{align}
	C^{\top}C \succcurlyeq e^{A^{\top}T}C^{\top}Ce^{AT},
\end{align}
then $f_{T,\sigma^2}(\Sigma)$ is a concave function of $\Sigma\succcurlyeq 0$.
\end{prop}
\begin{proof} 
Since $(A,C)$ is observable, $\cO$ is positive definite, and hence invertible. This implies that $\cW+\sigma^2\cO^{-1}$ is positive definite for every $\Sigma\succcurlyeq 0$. After some algebraic manipulations, we can rewrite $f_{T,\sigma^2}(\Sigma)$ as
\begin{align}
	f_{T,\sigma^2}(\Sigma) & = \frac{1}{2}\log_{2}\frac{\det(\sigma^2 \cO^{-1}+\cW)}{\det(\sigma^2\cO^{-1}+e^{AT}\cW e^{A^{\top}T})}\notag \\
	 &=\frac{1}{2}\log_{2}\frac{\det(\sigma^2\cO^{-1}+\cW)}{\det e^{2AT}\det(\sigma^2e^{-AT}\cO^{-1}e^{-A^{\top}T}+\cW)} \notag\\
	 &= \frac{1}{2}\log_{2} \det \Big[(\sigma^2\cO^{-1}+\cW)(\sigma^2e^{-AT}\cO^{-1}e^{-A^{\top}T}+\cW)^{-1}\Big] + \frac{1}{2}\log_2\det e^{-2AT} \notag\\
	 &= -\frac{1}{2}\log_{2} \det \Big[(\sigma^2\cO^{-1}+\cW)^{-1}(\sigma^2e^{-AT}\cO^{-1}e^{-A^{\top}T}+\cW)\Big] + \frac{1}{2}\log_2\det e^{-2AT} \notag\\
	 &= -\frac{1}{2 \ln 2}\ln \det\left(I+X^{-1}K\right) - \frac{T}{\ln 2}\tr(A),
\end{align}
where we have defined $X:=\sigma^2\cO^{-1}+\cW$ and $K:=\sigma^2e^{-AT}\cO^{-1}e^{-A^{\top}T}-\sigma^2 \cO^{-1}$.
Next, since 
\begin{enumerate}[(i)]
\item $X\succ 0$ is a linear function of $\Sigma$, and 
\item from \cite{DC01}, $\ln \det\left(I+X^{-1}K\right)$ is a convex function of $X$ if $K\succcurlyeq 0$,
\end{enumerate}
it follows that $f_{T,\sigma^2}(\Sigma)$ is concave if $K\succcurlyeq 0$. 
To conclude we notice that the latter condition is satisfied if 
\begin{align}
K\succcurlyeq 0 &\ \ \Leftrightarrow\ \   \sigma^2e^{-AT}\cO^{-1}e^{-A^{\top}T}- \sigma^2\cO^{-1}\succcurlyeq 0\notag\\
                &\ \ \Leftrightarrow\ \   e^{-AT}\cO^{-1}e^{-A^{\top}T}- \cO^{-1}\succcurlyeq 0\notag\\ 
                &\ \ \Leftrightarrow\ \ -e^{A^{\top}T}\cO e^{AT}+\cO\succcurlyeq 0\notag\\ 
                &\ \ \Leftrightarrow\ \  \int_0^T e^{A^\top t}(-e^{A^\top T}C^\top C e^{A T}+C^\top C)e^{AT} \text{d} t \succcurlyeq 0\notag\\
                &\ \ \Leftrightarrow\ \ -e^{A^\top T}C^\top C e^{A T}+C^\top C  \succcurlyeq 0.
\end{align}
This completes the proof.
\end{proof}

As a side remark, it is worth pointing out that the condition in \cref{prop:concave} always holds if $A$ is stable and normal and $C=I$, since $\|e^X\|=\lambda_{\max}(e^{X}(e^{X})^\top)<1$ for normal and stable $X$'s, where $\lambda_{\max}(\cdot)$ denotes the largest eigenvalue of a symmetric~matrix.

Next, we focus on the monotonicity properties of $f_{T,\sigma^2}(\Sigma)$ with respect to $\sigma^2$.
\begin{prop}[Monotonicity of $f_{T,\sigma^2}(\Sigma)$ w.r.t. $\sigma^2$] \label{prop:monotone} For all $\sigma^2_{2} \ge \sigma^2_{1} >0$ and $\Sigma\succcurlyeq 0$, it holds 
\begin{align}
	f_{T,\sigma^2_{1}}(\Sigma)&\ge f_{T,\sigma^2_{2}}(\Sigma),\label{eq:mono1}\\
	{\sigma^2_{1}}{f_{T,\sigma^2_{1}}(\Sigma)}&\le {\sigma^2_{2}}{f_{T,\sigma^2_{2}}(\Sigma)}.\label{eq:mono2}
\end{align}
\end{prop}
\begin{proof}
Given $\sigma^2>0$,  by using the properties of logarithms, $f_{T,\sigma^2}(\Sigma)$ can be rewritten as
\begin{align}
	f_{T,\sigma^2}(\Sigma) =  \sum_{i=1}^{n} \log_{2}\left( \frac{1+\lambda_{i}/\sigma^2}{1+\mu_{i}/\sigma^2}\right),
\end{align}
where $\{\lambda_{i}\}_{i=1}^{n}$, $\lambda_{1}\ge\dots\ge\lambda_{n}$, and $\{\mu_{i}\}_{i=1}^{n}$, $\mu_{1}\ge\dots\ge\mu_{n}$, are the eigenvalues of the positive (semi)definite matrices $\cO^{1/2}\cW\cO^{1/2}$ and $\cO^{1/2}(\cW-B\Sigma B^{\top})\cO^{1/2}$, respectively. By virtue of Weyl's Monotonicity Theorem \cite[Corollary III.2.3]{B13}, it holds $\lambda_{i}\ge \mu_{i}$ for all $i=1,2,\dots,n$. It is now a matter of direct computation to show that if $\lambda>\mu>0$,
\begin{align}
	\frac{\de}{\de x}\,  \log_{2}\left( \frac{1+\lambda/x}{1+\mu/x}\right)>0, \quad x>0,\\
	\frac{\de}{\de x}\,  x\log_{2}\left( \frac{1+\lambda/x}{1+\mu/x}\right)<0, \quad x>0.
\end{align}
In view of the latter inequalities, it follows that each term in the sum defining $f_{T,\sigma^2}(\Sigma)$ is a monotonically decreasing function of $\sigma^2>0$, whereas each term in the sum defining $\sigma^2 f_{T,\sigma^2}(\Sigma)$ is monotonically increasing function of $\sigma^2>0$.
\end{proof}

\begin{prop}[Monotonicity of $f_{T,\sigma^2}(\Sigma)$ w.r.t.~$T$] \label{prop:monotoneT} Suppose that the pair $(A,C)$ is observable and $\sigma^2>0$. For all $T_{1}>0$, $T_{2} =hT_{1}$, $h\in\N_{>0}$, and $\Sigma\succcurlyeq 0$, it holds 
 \begin{align}
 	f_{T_{2},\sigma^2}(\Sigma)\ge f_{T_{1},\sigma^2}(\Sigma).
 \end{align} 
 \end{prop}
 \begin{proof}
 Since $T_{2}\ge T_{1}$, we have
 \begin{align}\label{eq:OT1T2}
 \cO_{T_{2}}:=\int_0^{T_2} e^{A^\top t} C^\top C e^{At} \de t\succcurlyeq \int_0^{T_1} e^{A^\top t} C^\top C e^{At} \de t=:\cO_{T_{1}}\succ 0,
 \end{align}
 where positive definiteness follows from observability of the pair $(A,C)$. Furthermore, since $T_{2}= h T_{1}$, $h\in\N_{>0}$,
 \begin{align}\label{eq:WT1T2}
 	 \cW_{T_{1}}:=\sum_{k=0}^{\infty}e^{A kT_{1}}B\Sigma  B^{\top}e^{A ^{\top}kT_{1}} \succcurlyeq \sum_{k=0}^{\infty}e^{A k h T_{1}}B\Sigma  B^{\top}e^{A ^{\top}k hT_{1}}=: \cW_{T_{2}}.
 \end{align}
 Next, we can rewrite $f_{T,\sigma^2}(\Sigma)$ as
 \begin{align}\label{eq:monotone-T}
	 f_{T,\sigma^2}(\Sigma) &= \frac{1}{2} \log_2 \det \Big[\left(\sigma^2I+ \cO \cW \right) \left(\sigma^2 I+  \cO (\cW-B\Sigma B^\top) \right)^{-1}\Big]\notag\\
	 &= \frac{1}{2} \log_2 \det \Big[ I+ \cO B\Sigma B^\top\left(\sigma^2I+ \cO (\cW-B\Sigma B^\top) \right)^{-1}\Big]\notag\\
	 &= \frac{1}{2} \log_2 \det \Big[ I+ B\Sigma B^\top\left(\sigma^2I+ \cO (\cW-B\Sigma B^\top) \right)^{-1}\cO \Big]\notag\\
	 &= \frac{1}{2}\log_{2} \det \Big[I+B\Sigma B^\top((\cW-B\Sigma B^\top)+{\sigma^2\cO^{-1}})^{-1}\Big]\notag\\
	 &= \frac{1}{2}\log_{2} \det \Big[I+(B\Sigma B^\top)^{1/2}(e^{AT}\cW e^{A^{\top}T}+{\sigma^2 \cO^{-1}})^{-1}(B\Sigma B^\top)^{1/2}\Big],
\end{align}
In view of \Cref{eq:OT1T2,eq:WT1T2}, we have $\cO_{T_{2}}^{-1}\preccurlyeq \cO_{T_{1}}^{-1}$, and
\begin{align}
 &\cW_{T_1}+\sigma^2\cO_{T_1}^{-1} \succcurlyeq \cW_{T_2}+\sigma^2\cO_{T_2}^{-1}\notag\\
 \Rightarrow\ \ \ & \left(\cW_{T_1}+\sigma^2\cO_{T_1}^{-1}\right)^{-1} \preccurlyeq \left(\cW_{T_2}+\sigma^2\cO_{T_2}^{-1}\right)^{-1}\notag\\
 \Rightarrow \ \ \ & (B\Sigma B^\top)^{1/2}\left(\cW_{T_1}+\sigma^2\cO_{T_1}^{-1}\right)^{-1}(B\Sigma B^\top)^{1/2} \preccurlyeq (B\Sigma B^\top)^{1/2}\left(\cW_{T_2}+\sigma^2\cO_{T_2}^{-1}\right)^{-1}(B\Sigma B^\top)^{1/2}\notag\\
 \Rightarrow\ \ \ & I+(B\Sigma B^\top)^{1/2}\left(\cW_{T_1}+\sigma^2\cO_{T_1}^{-1}\right)^{-1}(B\Sigma B^\top)^{1/2} 
 \preccurlyeq I+(B\Sigma B^\top)^{1/2}\left(\cW_{T_2}+\sigma^2\cO_{T_2}^{-1}\right)^{-1}(B\Sigma B^\top)^{1/2}.
 \end{align} 
 Finally, from \Cref{eq:monotone-T}, the latter inequality implies $f_{T_{2},\sigma^2}(\Sigma)\ge f_{T_{1},\sigma^2}(\Sigma)$.
 \end{proof}

We note that, in the limit $T\to \infty$, $\cW \to B\Sigma B^\top$ and $\cO\to \int_0^T e^{A^\top t}C^\top C e^{At}\, \text{d} t=\cO_\infty$, which are both finite matrices since $A$ is stable. Hence, it follows that
\begin{align}
    \lim_{T\to \infty} f_{T,\sigma^2}(\Sigma) = \log_2 \det\left(I+ \frac{1}{\sigma^2}\cO_\infty B\Sigma B^\top\right),
\end{align}
which is finite. From \cref{prop:monotoneT}, this in turn implies that $f_{T,\sigma^2}(\Sigma)$ attains its maximum for $T\to \infty$.

To conclude, we establish a monotonicity property of the capacity with respect to a particular choice of the input and output matrices $B$ and $C$.

\begin{prop}[Monotonicity of $\cC_{T}$ w.r.t.~$B$ and $C$] \label{prop:monotoneBC} Assume that the pair $(A,C)$ is observable, and let
\begin{align}
	B_{1}&=\begin{bmatrix}I_{m_{1}} \\ 0\end{bmatrix}, \quad C_{1}=\begin{bmatrix}0 & I_{p_{1}}\end{bmatrix},\\
	B_{2}&=\begin{bmatrix}I_{m_{2}} \\ 0\end{bmatrix}, \quad C_{2}=\begin{bmatrix}0 & I_{p_{2}}\end{bmatrix},
\end{align}
be two pairs of input and output matrices. If $m_{2}\ge m_{1}$ and $p_{2}\ge p_{1}$, then 
\begin{align}
	\cC_{T}(B_{1},C_{1})\le \cC_{T}(B_{2},C_{2}),
\end{align} 
where we made explicit the dependence of $\cC_{T}$ on the input matrix $B$ and output matrix $C$.
\end{prop}
\begin{proof}
First, as in the proof of \cref{prop:monotoneT}, we have that \Cref{eq:max-func} can be rewritten as
\begin{align}\label{eq:monotone-T2}
	 f_{T,\sigma^2}(\Sigma) = \frac{1}{2}\log_{2} \det \Big[I+(B\Sigma B^\top)^{1/2}(e^{AT}\cW e^{A^{\top}T}+{\sigma^2 \cO^{-1}})^{-1}(B\Sigma B^\top)^{1/2}\Big],
\end{align}
Next, since $p_{2}\ge p_{1}$, we have 
\begin{align}
	 \cO_{1}:=\int_{0}^{T}e^{A^{\top} t}C_{1}^{\top}C_{1}e^{At}\,\de t \preccurlyeq \int_{0}^{T}e^{A^{\top} t}C_{2}^{\top}C_{2}e^{At}\,\de t =: \cO_{{2}}.
\end{align}
 From the previous inequality and the observability of $(A,C)$, we have $\cO_1^{-1}\succcurlyeq \cO_2^{-1}$. This in turn implies that 
 \begin{align}
 &\cW+\sigma^2\cO_1^{-1} \succcurlyeq \cW+\sigma^2\cO_2^{-1} \notag\\
 \Rightarrow\ \ \ & \left(\cW+\sigma^2\cO_1^{-1}\right)^{-1} \preccurlyeq \left(\cW+\sigma^2\cO_2^{-1}\right)^{-1} \notag\\
 \Rightarrow \ \ \ & (B\Sigma B^\top)^{1/2}\left(\cW+\sigma^2\cO_1^{-1}\right)^{-1}(B\Sigma B^\top)^{1/2} \preccurlyeq (B\Sigma B^\top)^{1/2}\left(\cW+\sigma^2\cO_2^{-1}\right)^{-1}(B\Sigma B^\top)^{1/2} \notag\\
 \Rightarrow\ \ \ & I-(B\Sigma B^\top)^{1/2}\left(\cW+\sigma^2\cO_1^{-1}\right)^{-1}(B\Sigma B^\top)^{1/2} 
 \preccurlyeq I-(B\Sigma B^\top)^{1/2}\left(\cW+\sigma^2\cO_2^{-1}\right)^{-1}(B\Sigma B^\top)^{1/2}
 \end{align} 
 which in view of \Cref{eq:monotone-T} yields \begin{align}f_{T,\sigma^2}(\Sigma,B,C_{2})\ge f_{T,\sigma^2}(\Sigma,B,C_{1}),\end{align} for any $B\in\R^{n\times m}$ and $\Sigma\succcurlyeq 0$, where we made explicit the dependence of $f_{T,\sigma^2}(\cdot)$ on the input matrix $B$ and output matrix $C$. From the previous inequality, it follows that
 \begin{align}\label{eq:monoC}
    \cC_{T}(B,C_{1}) = \max_{\Sigma \succcurlyeq 0, \tr\Sigma=P} f_{T,\sigma^2}(\Sigma,B,C_{1})\le \max_{\Sigma \succcurlyeq 0, \tr\Sigma=P} f_{T,\sigma^2}(\Sigma,B,C_{2})= \cC_{T}(B,C_{2}).
 \end{align}
 Now, let us define
\begin{align}
	\cS_{1}:=\{B_{1}\Sigma_{1}B_{1}\,:\, \Sigma_{1}\succcurlyeq 0, \ \tr\,\Sigma_{1}= P\}, \quad \cS_{2}:=\{B_{2}\Sigma_{2}B_{2}\,:\, \Sigma_{2}\succcurlyeq 0, \ \tr\,\Sigma_{2}=P\}.
\end{align}
Since $m_{2}\ge m_{1}$, we have $\cS_{1}\subseteq \cS_{2}$. This in turn implies that
\begin{align}\label{eq:monoB}
	\cC_{T}(B_{1},C)=\max_{\Sigma \succcurlyeq 0, \tr\Sigma=P} f_{T,\sigma^2}(\Sigma,B_{1},C) = \max_{\Sigma \in \cS_1} f_{T,\sigma^2}(\Sigma,I,C) \le \max_{\Sigma \in \cS_2} f_{T,\sigma^2}(\Sigma,I,C)= \max_{\Sigma \succcurlyeq 0, \tr\Sigma=P} f_{T,\sigma^2}(\Sigma,B_{2},C) =\cC_{T}(B_{2},C).
\end{align}
Finally, a combination of \Cref{eq:monoC,eq:monoB} yields the thesis.
\end{proof}

\cref{prop:monotoneBC} in particular implies that the maximum capacity is always attained by picking $B=C=I$, that is, by selecting all nodes in the network as input/output nodes.

\section{Information rate in the low and high noise regime}

Here, we analyze the behavior of the information rate $\cR_T$ in the limit cases where $\sigma^2$ tends to either zero or infinity. 

\begin{thm}[Behavior of $\cR_{T}$ in the low noise regime]\label{prop:asymp-SNR-low}
Let $A \in\mathbb{R}^{n\times n}$ be a stable matrix and assume that the pair $(A,C)$ is observable. For all $T>0$, it holds
\begin{align}
\lim_{\sigma^2\to 0}\cR_{T} &= -\frac{1}{\ln 2}\tr(A). \label{eq:Rsigma}
\end{align}
Further, as $\sigma^2\to 0$, the optimal $\Sigma$ in \Cref{eq:cap} is any positive semidefinite unit-trace matrix such that $(A,B\Sigma^{1/2})$ is controllable.
\end{thm}
\begin{proof}
Consider $f_{T,\sigma^2}(\Sigma)$ as defined in \Cref{eq:max-func}. Since $f_{T,\sigma^2}(\Sigma)$ and $\smash{\displaystyle\max_{\Sigma\succcurlyeq 0, \tr\Sigma =1}f_{T,\sigma^2}(\Sigma)}$ are monotonically decreasing functions of $\sigma^2$ (\cref{prop:monotone}, \Cref{eq:mono1}), we have that 
\begin{align}
	 \lim_{\sigma^2\to 0} \max_{\Sigma\succcurlyeq 0, \tr\Sigma =1}f_{T,\sigma^2}(\Sigma) =  \sup_{\sigma^2\ge 0} \max_{\Sigma\succcurlyeq 0, \tr\Sigma =1}f_{T,\sigma^2}(\Sigma) =  \max_{\Sigma\succcurlyeq 0, \tr\Sigma =1} \sup_{\sigma^2\ge 0}  f_{T,\sigma^2}(\Sigma) = \max_{\Sigma\succcurlyeq 0, \tr\Sigma =1}\lim_{\sigma^2\to 0}f_{T,\sigma^2}(\Sigma).
\end{align}

If $\Sigma$ maximizing $\displaystyle\lim_{\sigma^2\to 0}f_{T,\sigma^2}(\Sigma)$ is such that $(e^{AT},B\Sigma^{1/2})$ is a controllable pair, then $\cW>0$, and we have
\begin{align}
	\frac{1}{T}\lim_{\sigma^2\to 0} f_{T,\sigma^2}(\Sigma)&=\frac{1}{2T}\log_{2}\frac{\det(\cO\cW)}{\det(\cO e^{A T}\cW e^{A^{\top}T})}\notag\\
	&=-\frac{1}{2T}\log_{2}\det(e^{A T})^{2}=-\frac{1}{T\ln 2}\ln\det(e^{A T})\notag\\
	&=-\frac{1}{T\ln 2}\tr(A T)=-\frac{1}{\ln 2}\tr(A)\label{eq:proof-snr-limit},
\end{align}
where in the second equation we used the fact that $\cO$ is invertible, in view of the observability of $(A,C)$. If the maximizing $\Sigma$ is such that $(e^{AT},B\Sigma^{1/2})$ is \emph{not} controllable, then, via a suitable similarity transformation $Q$, we can bring $e^{AT}$ and $B\Sigma^{1/2}$ in Kalman canonical form \cite[Chapter 16]{H09}, namely
\begin{align}
	\overline{A}:=Q^{-1}e^{AT}Q=\begin{bmatrix} A_{11} & A_{12} \\ 0 & A_{22}\end{bmatrix}, \quad \overline{B}:=Q^{-1}B\Sigma^{1/2}=\begin{bmatrix} B_{1}\\ 0 \end{bmatrix},
\end{align} 
where $(A_{11},B_{1})$ forms a controllable pair. By partitioning $\cO$ conformably to the block partition of $\overline{A}$, namely,
\begin{align}
	\overline{\cO}: = Q^{-\top}\cO Q^{-1}=\begin{bmatrix} \cO_{11} & \cO_{12} \\ \cO_{21} & \cO_{22}\end{bmatrix},\quad \cO_{11}>0,
\end{align}
it follows that, 
\begin{align}
	\frac{1}{T}\lim_{\sigma^2\to 0} f_{T,\sigma^2}(\Sigma)&=-\frac{1}{T}\log_{2}\det(A_{11}) < -\frac{1}{T}\log_{2}\det(e^{A T}),
\end{align}
since the eigenvalues of $A_{11}$ are a subset of eigenvalues of $e^{AT}$. This in turn implies that the maximum of $\lim_{\sigma^2\to 0}f_{T,\sigma^2}(\Sigma)$ cannot be achieved by any $\Sigma$ that renders $(e^{AT},B\Sigma^{1/2})$ not controllable and, therefore, its value is given by \Cref{eq:proof-snr-limit}.
\end{proof}

\begin{thm}[Behavior of $\cR_{T}$ in the high noise regime]\label{prop:asymp-SNR-high}
Let $A \in\mathbb{R}^{n\times n}$ be a stable matrix. For all $T>0$, it holds
\begin{align}
\lim_{\sigma^2\to \infty} \sigma^2\cR_{T} &= \frac{1}{2\ln 2 T} \|B \cO B^\top\|. \label{eq:Csigma}
\end{align}
Further, as $\sigma^2\to \infty$, the optimal $\Sigma$ in \Cref{eq:cap} is given by $\Sigma^\star = vv^\top/(v^\top v)$, where $v\in\R^m$ is the eigenvector corresponding to the maximum eigenvalue of $B \cO B^\top$.
\end{thm}
\begin{proof}
Since $\sigma^2 f_{T,\sigma^2}(\Sigma)$ and $\smash{\displaystyle\max_{\Sigma \succcurlyeq 0, \tr\Sigma =1}\sigma^2 f_{T,\sigma^2}(\Sigma)}$ are monotonically increasing functions of $\sigma^2$ (\cref{prop:monotone}, \Cref{eq:mono2}),
\begin{align}\label{eq:exchanging-lim-max-2}
	& \lim_{\sigma^2\to \infty} \max_{\Sigma \succcurlyeq 0, \tr\Sigma =1}\sigma^2 f_{T,\sigma^2}(\Sigma) =  \sup_{\sigma^2\ge 0} \max_{\Sigma \succcurlyeq 0, \tr\Sigma =1}\sigma^2 f_{T,\sigma^2}(\Sigma) = \max_{\Sigma \succcurlyeq 0, \tr\Sigma =1} \sup_{\sigma^2\ge 0}  \sigma^2 f_{T,\sigma^2}(\Sigma) = \max_{\Sigma \succcurlyeq 0, \tr\Sigma =1}\lim_{\sigma^2\to \infty}\sigma^2 f_{T,\sigma^2}(\Sigma).
\end{align}
Using the Taylor expansion of the natural logarithm $\ln(1+x)=x+\text{h.o.t.}$ as $x\to 0$, it holds
\begin{align}
	\ln\det(I +\cO\cW/\sigma^2)&=\tr(\cO\cW)/\sigma^2+\text{h.o.t.}\quad \text{ as }\ \ \sigma^2\to \infty, \\
	 \ln\det(I +\cO(\cW-B^\top \Sigma B)/\sigma^2)&= \tr(\cO(\cW-B^\top \Sigma B))/\sigma^2+\text{h.o.t.}\quad \text{ as }  \ \ \sigma^2\to \infty,
\end{align}
so that
\begin{align}
\lim_{\sigma^2\to \infty} \sigma^2 f_{T,\sigma^2}(\Sigma)&=\frac{1}{2\ln 2}\lim_{\sigma^2\to \infty} \sigma^2\ln\frac{\det(I +\cO\cW/\sigma^2)}{\det(I+\cO(\cW-B^\top \Sigma B)/\sigma^2)} \notag\\
&=\frac{1}{2\ln 2}\lim_{\sigma^2\to \infty} \sigma^2\left(\ln\det(I +\cO\cW/\sigma^2) - \ln\det(I+\cO(\cW-B^\top \Sigma B)/\sigma^2)\right) \notag\\
&=\frac{1}{2\ln 2}  \tr(\cO B^\top \Sigma B).
\end{align}
Next, we note that the following inequality holds \cite{fang1994inequalities}
\begin{align}
	\tr(\cO B^\top \Sigma B) = \tr(B \cO B^\top \Sigma) \le \|B \cO B^\top\|\tr(\Sigma).
\end{align}
On the other hand, by picking $\Sigma^{\star}= vv^\top/(v^\top v)$, with $v\in\R^n$ being the eigenvector corresponding to the maximum eigenvalue of $B \cO B^\top$, we have $\tr(B \cO B^\top\Sigma^\star) = \|B \cO B^\top\|\tr(\Sigma^\star) = \|B \cO B^\top\|$. Hence,
\begin{align}
	\max_{\Sigma \succcurlyeq 0, \tr\Sigma =1} \tr(B \cO B^\top\Sigma) =  \|B \cO B^\top\|,
\end{align}
from which \Cref{eq:Csigma} follows using \Cref{eq:exchanging-lim-max-2}.
\end{proof}

\section{Information rate of normal networks}

We derive here an explicit expression for the information information rate of networks described by a normal adjacency matrix with $B=C=I$. To this end, we first establish an instrumental lemma. In what follows, $\diag(X)$ will denote the diagonal matrix with the diagonal entries of $X$, if $X$ is a matrix, or the entries of $X$ along the diagonal, if $X$ is a vector.

\begin{lem}\label{lemma:pos-def}
Let $P\in\C^{n\times n}$ be a positive definite Hermitian matrix. It holds
$
	 \diag(P^{-1})\succcurlyeq (\diag\, P)^{-1}.
$
\end{lem}

\begin{proof}
Let $\{e_{i}\}_{i=1}^{n}$ be the canonical basis in $\R^{n}$, by Cauchy--Schwarz inequality 
$
	(e_{i}^{\top}P e_{i})(e_{i}^{\top}P^{-1} e_{i})=\| P^{1/2}e_{i}\|^{2}\|P^{-1/2}e_{i}\|^{2}\ge  (e_{i}^{\top}P^{1/2}P^{-1/2}e_{i})^{2}=1,
$
so that $e_{i}^{\top}P^{-1} e_{i}\ge (e_{i}^{\top}P e_{i})^{-1}$ for all $i=1,\dots,n$.
\end{proof}

\begin{thm}[information rate of normal networks]\label{prop:normalcap}
If $B=C=I$ and $A \in\R^{n\times n}$ is a normal and stable matrix with eigenvalues $\{\lambda_{i}\}_{i=1}^{n}$, then 
\begin{align}\label{eq:Cnormal}
\cR_{T} = \max_{\substack{\{P_{i}\}_{i=1}^{n}, P_{i}\ge 0 \\ \text{\ s.t.\ }\sum_{i=1}^{n} P_{i} = P}}\sum_{i=1}^{n} \cR_{T}(P_{i},\lambda_{i}),
\end{align} 
with
\begin{align}\label{eq:Capnormi}
\cR_{T}(P,\lambda):=\frac{1}{2T}\log_{2}\frac{\frac{P}{\sigma^2}-2\mathrm{Re}\,\lambda}{\frac{P}{\sigma^2}\,e^{2T\mathrm{Re}\,\lambda}-2\mathrm{Re}\,\lambda}.
\end{align} 
\end{thm}
\begin{proof}
Consider the function $f_{T,\sigma^2}$ as previously defined in \Cref{eq:max-func}.
We will show that the matrix $\Sigma \succcurlyeq 0$, $\tr(\Sigma)=1$, maximizing $f_{T,\sigma^2}(\Sigma)$ is diagonal w.r.t.~the same basis diagonalizing $A$. Along the lines of the proof of \cref{prop:concave}, we first note that $f_{T,\sigma^2}$ can be equivalently rewritten as
\begin{align}\label{eq:proof-normal-1}
f_{T,\sigma^2}(\Sigma) = \frac{1}{2\ln 2}\ln \det\left(I-\sigma^2 KX^{-1}\right)-\frac{T}{\ln 2}\tr(A),
\end{align}
where $X:=\cW+\sigma^2e^{-AT}\cO^{-1}e^{-A^{\top}T}$ and $K:=e^{-AT}\cO^{-1}e^{-A^{\top}T}-\cO^{-1}$. Next, if $A$ is stable and normal, it can be unitarily diagonalized in the form $A=-U^{\dagger}\Lambda U$, where $U$ is a unitary matrix and $\Lambda$ is diagonal with diagonal entries $\{\lambda_{i}\}_{i=1}^{n}$ having positive real part. Since $C=I$, we can rewrite $\cO$ as
\begin{align}
	\cO= U^{\dagger}\left[\frac{1}{2}\mathrm{Re}(\Lambda)^{-1}\left(I-e^{-2\mathrm{Re}(\Lambda)T}\right)\right]U,
\end{align} 
where $\mathrm{Re}(\cdot)$ denotes element-wise real part when applied to matrices, so that \Cref{eq:proof-normal-1} can be rewritten as 
\begin{align}
f_{T,\sigma^2}(\Sigma) = \frac{1}{2\ln 2}\ln \det\left(I-\sigma^2\overline{K}^{1/2}\overline{X}^{-1}\overline{K}^{1/2}\right) -\frac{T}{\ln 2}\tr(A),
\end{align}
where $\overline{X}:=U\cW U^{\dagger}+\frac{\sigma^2}{2}\mathrm{Re}(\Lambda)\left(e^{2\mathrm{Re}(\Lambda)T}-I\right)^{-1}$ and $\overline{K}:=\frac{1}{2}\mathrm{Re}(\Lambda)e^{2\mathrm{Re}(\Lambda)T}\succ 0$. Now assume, by contradiction, that the optimal $\Sigma^{\star}$, $\tr(\Sigma^{\star})=1$, maximizing $f_{T,\sigma^2}(\Sigma)$  yields a matrix $\bar\Sigma^{\star}:=U\Sigma^{\star}U^{\dagger}$ which is not diagonal. This implies that $Z^{\star}:=\overline{K}^{1/2}\overline{X}^{-1}\overline{K}^{1/2}$ is not diagonal, since
\begin{enumerate}[(i)]
    \item  $U\cW U^{\dagger}=\sum_{k=0}^{\infty}e^{-\Lambda kT} \bar{\Sigma} e^{-\Lambda^{\dagger}kT}$ is not diagonal,
    \item $\frac{\sigma^2}{2}\mathrm{Re}(\Lambda)\left(e^{2\mathrm{Re}(\Lambda)T}-I\right)^{-1}$ and $\overline{K}$ are diagonal matrices, and
    \item the inverse of a non-singular, non-diagonal matrix cannot be diagonal.
\end{enumerate}
Let $\bar\Sigma_{d}:=\diag(\bar\Sigma^{\star})$ and consider the corresponding diagonal matrix $Z_{d}:=K^{1/2}\overline{X}_{d}^{-1}K^{1/2}$, where \begin{align}\overline{X}_{d}:=U\cW_{d} U^{\dagger}+\frac{\sigma^2}{2}\mathrm{Re}(\Lambda)\left(e^{2\mathrm{Re}(\Lambda)T}-I\right)^{-1},\end{align} and $\cW_{d}:=\sum_{k=0}^{\infty}e^{-\Lambda kT} \bar{\Sigma}_{d} e^{-\Lambda^{\dagger}kT}$. Since for every positive definite matrix $P\succ 0$ it holds $\diag(P^{-1})\succcurlyeq (\diag\, P)^{-1}$ (\cref{lemma:pos-def}), we have that
\begin{align}\label{eq:proof-normal-2}
	\diag(Z^{\star})=\overline{K}^{1/2}\diag(\overline{X}^{-1})\overline{K}^{1/2}\succcurlyeq \overline{K}^{1/2}\bar{X_{d}}^{-1}\overline{K}^{1/2} =Z_{d}.
\end{align}
Eventually, we argue that
\begin{align}
	\det(I-Z^{\star}) &< \det(I-\diag(Z^{\star})) \overset{\cref{eq:proof-normal-2}}{\le} \det(I-Z_{d}),\label{eq:proof-normal-3}
\end{align}
where the first inequality follows from the fact that $I-Z^{\star}\succ 0$ and from Hadamard's inequality $\det P\le \det\diag(P)$, for a positive definite $P\succ 0$ (with equality attained if and only if $P$ is diagonal, cf. \cite[p.~505]{HJ12}). The inequality in \Cref{eq:proof-normal-3} contradicts the optimality of $\Sigma^{\star}$ and therefore the matrix $U\Sigma^{\star}U^{\dagger}$ must be diagonal. In view of this fact, the desired expression in \Cref{eq:Cnormal} follows by direct computation.
\end{proof}

From the above result we have the following interesting corollary.

\begin{cor}[Optimal power allocation for normal networks]
If $B=C = I$ and $A$ is normal, then $\cR_{\max} := \max_{T\ge 0}\cR_{T}$ is achieved for $T\to 0$ and the optimal power allocation is
\begin{align}
	P_{i}= \frac{\mathrm{Re}\,\lambda_{i}}{\tr A} P,
\end{align}
which yields
\begin{align}\label{eq:Rmaxnormal}
	\cR_{\max} = \frac{1}{\ln 2} \frac{P \tr A}{2\sigma^2 \tr A-P}.
\end{align}
\end{cor}
\begin{proof}
Since $\cR_{T}$ is the sum of $n$ terms of the form in \Cref{eq:Capnormi} and each of these terms is a monotonically decreasing function of $T$, it follows that the maximum of $\cR_{T}$ is achieved for $T\to 0$. For $T\to 0$, the expression of $\cR_{T}$ becomes
\begin{align}\label{eq:Rnormal-cor}
\cR_{T} = \frac{1}{\ln 2}\max_{\substack{\{P_{i}\}_{i=1}^{n}, P_{i}\ge 0 \\ \text{\ s.t.\ }\sum_{i=1}^{n} P_{i} = P}}\sum_{i=1}^{n} \frac{\frac{P_{i}}{\sigma^{2}}\mathrm{Re}\,\lambda_{i}}{2\mathrm{Re}\,\lambda_{i}-\frac{P_{i}}{\sigma^{2}}}.
\end{align}  
Consider the Lagrangian of the above constrained optimization problem
\begin{align}
	\cL(P_1,\dots,P_n,\mu)=\sum_{i=1}^{n} \frac{\frac{P_{i}}{\sigma^{2}}\mathrm{Re}\,\lambda_{i}}{2\mathrm{Re}\,\lambda_{i}-\frac{P_{i}}{\sigma^{2}}} + \mu\left(P-\sum_{i=1}^{n}P_{i}\right),
\end{align}
where $\mu\in\R$ is the Lagrangian multiplier. By equating $\partial \cL /\partial P_{i}$ to zero, we have
\begin{align}
	\frac{\partial \cL}{\partial P_{i}} = \frac{2(\mathrm{Re}\,\lambda_{i})^{2}}{\left(-2\mathrm{Re}\,\lambda_{i}+\frac{P_{i}}{\sigma^{2}}\right)^{2}}-\mu=0 \Rightarrow \frac{P_{i}}{\sigma^{2}}=\left(2\pm\sqrt{\frac{2}{\mu}}\right)\mathrm{Re}\,\lambda_{i}, \quad i=1,2,\dots,n.
\end{align}
In view of the latter equation,
\begin{align}
	\frac{P}{\sigma^{2}}=\sum_{i=1}^{n}\frac{P_{i}}{\sigma^{2}}=\left(2\pm\sqrt{\frac{2}{\mu}}\right)\sum_{i=1}^{n} \mathrm{Re}\,\lambda_{i} = \left(2\pm\sqrt{\frac{2}{\mu}}\right)\tr A\Rightarrow \frac{P_{i}}{\sigma^{2}}=\frac{P}{\tr A}\mathrm{Re}\,\lambda_{i}.
\end{align} 
Eventually, by substituting the above optimal signal-to-noise ratios $P_{i}/\sigma^{2}$ into \Cref{eq:Rnormal-cor}, we obtain
\begin{align}
	\cR_{\max} = \frac{1}{\ln 2} \frac{P \tr A}{2\sigma^2 \tr A-P}.
\end{align}
which ends the proof.
\end{proof}
It is worth pointing out that, from \cref{prop:monotoneBC}, \Cref{eq:Cnormal} provides an upper bound to the achievable rate of any normal network for \emph{any} choice of input and output node subsets.  Furthermore, the expression of $\cR_{\max}$ in \Cref{eq:Rmaxnormal} corresponds to the maximum information rate that can be attained by a network described by a normal matrix $A$ for a fixed noise~level.


\section{Measures of matrix non-normality and network indicators of non-normality}

Mathematically speaking, matrix non-normality is not easy to measure and there does not exist a unique scalar parameter that quantifies it. However, several scalar estimates of matrix non-normality have been proposed and studied in the literature, see \cite{EP87}, or Chapter 48 of \cite{Trefethen2005} for a complete survey. There are two general approaches for quantifying the ``degree'' of non-normality of a matrix $A\in\mathbb{R}^{n\times n}$. The first and classic approach is to measure non-normality in terms of departure from some algebraic properties characterizing the set of normal matrices. In particular, metrics that fall into this category are zero when evaluated at normal matrices. Examples include:
\begin{itemize}
    \item the distance from commutativity, $c(A) = \| AA^\top - A^\top A\|_F$, where $\|\cdot\|_F$ is the Frobenius norm;
    \item the Henrici's departure from normality, $h(A) = ( \|A\|_F^2 - \sum_{z\in\lambda(A)} |z|^2)^{1/2}$, where $\lambda(\cdot)$ denotes the set of eigenvalues (or spectrum) of a matrix.
\end{itemize}
Alternatively, one can consider the linear dynamical system governed by $A$, and quantify non-normality in terms of the ability of this system to transiently amplify its trajectories. This ability is, in turn, related to the sensitivity of the spectrum of $A$ to perturbations of the entries of $A$ \cite{Trefethen2005}.
Some standard measures of non-normality that carry such a ``dynamical'' interpretation~are:
\begin{itemize}
    \item the numerical abscissa, $\omega(A) = \max_{z\in\lambda((A+A^\top)/2)} \ z$;
    \item the condition number of the eigenvector matrix of $A$, $\kappa(A)=\|V\|\|V^{-1}\|$, where the columns of $V$ are the eigenvectors~of~$A$.\footnote{ Here, we assume that $A$ is diagonalizable. Note that, if $A$ is normal, then $\kappa(A)=1$.}
\end{itemize}

\begin{figure}[htbp]
    \centering
    \includegraphics{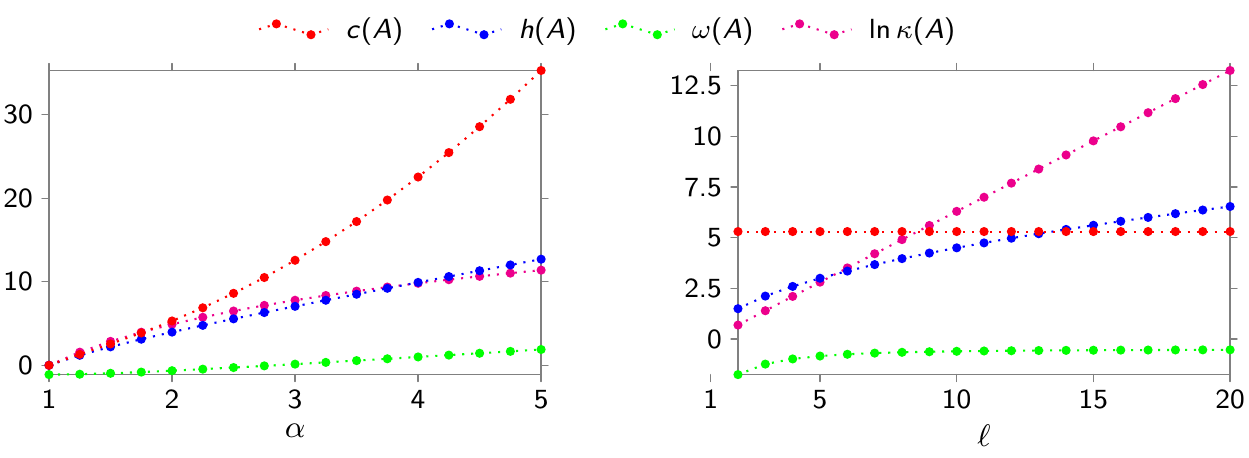}
    \caption{\textbf{Non-normality metrics and parameters of chain network.} Correlation between standard non-normality metrics and parameters $\alpha$ and $\ell$ of the chain network described in Equation (9) of the main text with $\gamma=-3$ and $\beta=1$. For better visualization, $\kappa(A)$ is plotted in logarithmic scale. Left plot: $\ell=n=8$. Right plot: $\ell=n$, $\alpha=2$.}
    \label{fig:nn_degree_chain}
\end{figure}

\begin{figure}[htbp]
    \centering
    \includegraphics{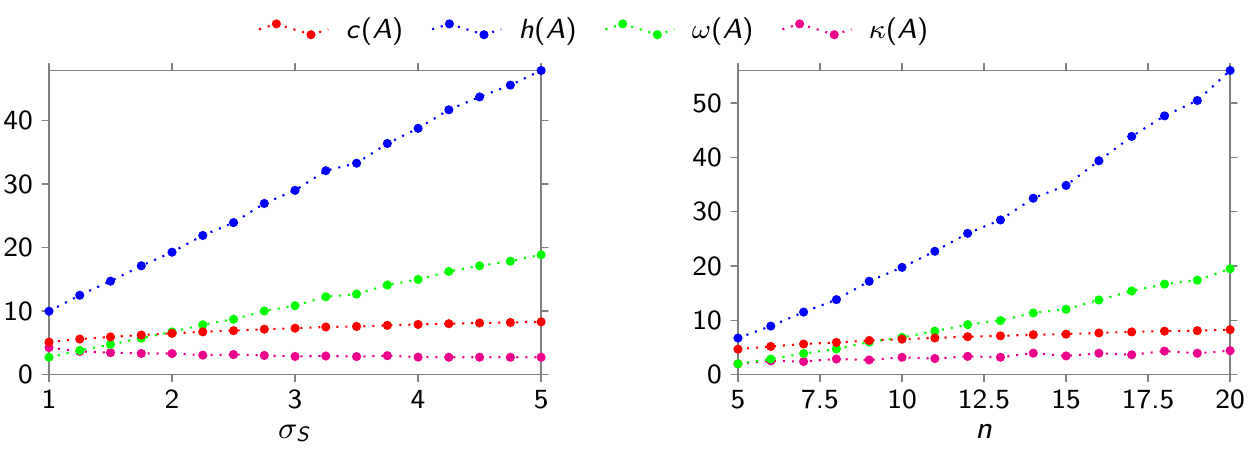}
    \caption{\textbf{Non-normality metrics and parameters of heterogeneous network.} Correlation between standard non-normality metrics and parameter $\sigma_S$ and dimension $n$ of heterogenous topologies. The curves are the average over $500$ realizations of the random model in Equation (11) of the main text with scale parameter $\omega^{-2}=\nu-n-1$ and $\nu=24+n$ degrees of freedom. Left plot: $n=10$. Right plot:~$\sigma_S=2$.}
    \label{fig:nn_degree_random}
\end{figure}

In the network considered in our paper, the above-listed non-normality metrics correlate with some scalar network parameters. These parameters can thus be thought of as knobs regulating the ``degree'' of non-normality of the network. Specifically, for the chain network in Equation (9) of the main text, as $\alpha$ (directionality strength) grows, then all of the above-listed measures of non-normality increase as well (\Cref{fig:nn_degree_chain}, left plot). For $\alpha\ne 1$, $h(A)$, $\kappa(A)$, and $\omega(A)$ are positively correlated with the chain length $\ell$, whereas $c(A)$ is independent of $\ell$ (\Cref{fig:nn_degree_chain}, right plot). A similar correlation behavior is found when considering the more general layered topologies described in Supplementary Note 6 (not shown). Finally, for heterogeneous topologies generated via Equation (11) of the main text, the parameter $\sigma_S$ (st.~dev.~of the entries of the skew-symmetric matrix $S$) and the network dimension $n$ are highly correlated with the non-normality metrics $h(A)$ and $\omega(A)$, whereas they do not exhibit a clear correlation w.r.t. $c(A)$ and $\kappa(A)$ (\Cref{fig:nn_degree_random}).

If matrix $A$ represents the adjacency matrix of a network, then it can be shown that non-normality is related to (i) absence of cycles in the network, (ii) low reciprocity of directed edges, and (iii) presence of hierarchical organization (see \cite{ALC18} for further details). With reference to our examples of chain and layered topologies, these networks are clearly acyclic, with reciprocity inversely proportional to $\alpha$, and number of ``hierarchical levels'' equal to $\ell$.

Finally, we mention that, although for general networks the presence of long directed paths does not necessarily imply (strong) non-normality\footnote{ An example is a directed ring topology, which is always normal regardless of the directionality strength.}, if the network is stable (a standing assumption in our paper) the connection between strong non-normality (here interpreted in a dynamical sense) and presence of strongly directional network paths can be made more precise. We next provide a simple example of this fact by illustrating what happens to a directed chain in which we add one edge to form a directed ring, while we refer to \cite{BZ18} for a more rigorous characterization of the aforementioned connection for general networks with non-negative weights (self-loops excluded). In the example, we consider as non-normality measure the numerical abscissa.


Consider the directed chain described by the following adjacency matrix
\begin{align}\label{eq:line}
	A=\begin{bmatrix}
	\gamma       & 0     &   0       &          & \\ 
	\alpha& \gamma &0&\ddots & \\ 
	&\ddots& \ddots& \ddots & 0  \\ 
	& &\ddots& \gamma& 0 \\
	& && \alpha& \gamma 
	\end{bmatrix}\in\R^{n\times n},
\end{align}
with $\gamma\in\mathbb{R}$ and $\alpha>0$.
It can be shown that the numerical abscissa is
$$\omega(A)=\gamma+\alpha\cos\left(\frac{\pi}{n+1}\right)$$ 
which, for large $n$, is approximately equal to
$$\omega(A)\approx \gamma+\alpha.$$
The system described by $A$ exhibits transient amplification if and only if $\omega(A)>0$ \cite{Trefethen2005}. If $A$ is stable, then $\gamma<0$ and the previous condition implies that $\alpha$ should be sufficiently large (strong directionality) for the network to feature a high ``degree'' of non-normality.
If we add an edge so that the directed chain becomes a directed ring, matrix $A$ becomes
\begin{align}\label{eq:ring}
	A'=\begin{bmatrix}
	\gamma       & 0     &   0       &          &\alpha \\ 
	\alpha& \gamma &0&\ddots & \\ 
	&\ddots& \ddots& \ddots & 0  \\ 
	& &\ddots& \gamma& 0 \\ 
	& && \alpha& \gamma 
	\end{bmatrix},
\end{align}
which is indeed a normal matrix with the same trace of $A$. However, when $\alpha+\gamma>0$ the largest eigenvalue of $A'$ (which equals its numerical abscissa) is unstable. Thus in the case of strong directionality, the matrix $A'$ becomes unstable, so that the stability assumption is violated.

\section{Information rate of a class of non-normal networks}

Let $S\in\R^{n\times n}$ be the adjacency matrix of a weighted graph $\mathcal{G}=(\mathcal{V},\mathcal{E})$, where $\mathcal{V}$ and $\mathcal{E}$ denote the set of nodes and edges of $\mathcal{G}$, respectively, and consider the following matrix
\begin{align}\label{eq:A non-normal}
	A:=DSD^{-1},
\end{align}
where $D:=\diag(d_{1},d_{2},\dots,d_{n})$, $d_{i}>0$, $i=1,2,\dots,n$. Note that the graph with adjacency matrix $A$ has the same connectivity of $\mathcal{G}$ but, in general, different weights. Additionally, $A$ has the same eigenvalues of $S$, but, typically, a different ``degree'' of non-normality. The following result characterizes the information rate of the network with adjacency matrix $A$ as in \Cref{eq:A non-normal}.

\begin{thm}[information rate of networks with adjacency matrix $A=DS D^{-1}$]\label{thm:DAD}
Let $S\in\R^{n\times n}$ be the stable adjacency matrix of a weighted graph $\mathcal{G}=(\mathcal{V},\mathcal{E})$, and consider $A := D S D^{-1}$, where $D_1:=\diag(d_{1},d_{2},\dots,d_{n})$, $d_{i}>0$, $i=1,2,\dots,n$, is a positive diagonal matrix. 
Let $B=[e_{k_1},\dots,e_{k_m}]$ and $C=[e_{t_1},\dots,e_{t_p}]^\top$ be the input matrix and the output matrix respectively and let
$d_{\min}:=\min \{d_{k_1},\dots,d_{k_m}\}$ and $d_{\max}=\max \{d_{t_1},\dots,d_{t_p}\}$.
Then,
\begin{align}
    \cR_T(A,\sigma^2)\ge  \cR_T\bigg(S,\bigg(\frac{d_{\min}}{d_{\max}}\sigma\bigg)^2\bigg), \label{eq:d and noise}
\end{align}
where we made explicit the dependence of $\cR_T$ on the network adjacency matrix and noise variance. 
\end{thm}

\begin{proof} Let $r:=\arg\min \{d_{k_1},\dots,d_{k_m}\}$ and 
$q:=\arg\max \{d_{t_1},\dots,d_{t_p}\}$. Consider first the special case where $C=e_q$ and $B=e_r$. In view of the definition of $A$, we have $D e_q^\top e_q D=d_{\max}^2 e_q^\top e_q$. This in turn implies that
\begin{align}\label{eq:Ohat}
    \cO = \int_0^T e^{A^\top t} C^\top C e^{At} \de t = \int_0^T D^{-1} e^{S^\top t} D e_q^\top e_q D e^{St} D^{-1}\de t  = d_{\max}^2 D^{-1} \underbrace{\int_0^T  e^{S^\top t} e_q^\top e_q e^{St} \de t}_{=:\widehat{\cO}} D^{-1}.
\end{align}
Similarly, we have $D^{-1} e_r e_r^\top  D^{-1} = \frac{1}{d_{\min}^2}e_r e_r^\top$, so that we can rewrite $\cW$ as
\begin{align}\label{eq:What}
    \cW = \sum_{k=0}^{\infty}e^{A kT}B\Sigma B^\top e^{A ^{\top}kT} = \sum_{k=0}^{\infty}D e^{S kT} D^{-1} e_r e_r^\top  D^{-1} e^{S ^{\top}kT} D = \frac{1}{d_{\min}^2} D \underbrace{\sum_{k=0}^{\infty} e^{S kT}e_r e_r^\top  e^{S ^{\top}kT}}_{=:\widehat{\cW}} D.
\end{align}
By substituting \Cref{eq:Ohat,eq:What} in the expression of $\cR_T$, we have
\begin{align}
\cR_T(A,\sigma^2)&= \frac{1}{2T}\,\max_{\substack{\Sigma\succcurlyeq 0,\, \tr\Sigma  = P}} \log_{2} \frac{\det(\sigma^{2}I+\cO \cW)}{\det(\sigma^{2}I+\cO(\cW -B\Sigma B^{\top} ))}= \frac{1}{2T}\,\max_{\substack{\Sigma\succcurlyeq 0,\, \tr\Sigma  = P}} \log_{2} \frac{\det(\sigma^{2}I+\cO \cW)}{\det(\sigma^{2}I+\cO e^{AT}\cW e^{A^\top T})}\notag\\
&= \frac{1}{2T}\,\max_{\substack{\Sigma\succcurlyeq 0,\, \tr\Sigma  = P}} \log_{2} \frac{\det(\sigma^{2}I+\left(\frac{d_{\max}}{d_{\min}}\right)^2 D^{-1}\widehat{\cO} D^{-1} D \widehat{\cW} D)}{\det(\sigma^{2}I+\left(\frac{d_{\max}}{d_{\min}}\right)^2 D^{-1}\widehat{\cO} D^{-1} De^{ST} D^{-1}D \widehat{\cW} D D^{-1}e^{S^\top T} D)}\notag\\
&= \frac{1}{2T}\,\max_{\substack{\Sigma\succcurlyeq 0, \, \tr\Sigma  = P}} \log_{2} \frac{\det\bigg(\left(\frac{d_{\min}}{d_{\max}}\sigma\right)^2 I+ \widehat{\cO}  \widehat{\cW} \bigg)}{\det\bigg(\left(\frac{d_{\min}}{d_{\max}}\sigma\right)^2 I+ \widehat{\cO} e^{ST} \widehat{\cW}  e^{S^\top T}\bigg)}= \cR_T\bigg(S,\bigg(\frac{d_{\min}}{d_{\max}}\sigma\bigg)^2\bigg).
\end{align}
Finally, for the general case where $B=[e_{k_1},\dots,e_{k_m}]$ is such that $r \in \{k_1,\dots,k_m\}$, and $C=[e_{t_1},\dots,e_{t_p}]^\top$ is such that $q \in \{t_1,\dots,t_p\}$, the inequality in \Cref{eq:alpha} directly follows from \cref{prop:monotoneBC}.
\end{proof}

\cref{thm:DAD} in particular implies that, as the heterogeneity of the elements of $D$ grows unbounded, we recover the same value of information rate of the \emph{noiseless} case. 
\begin{cor}[Limiting behavior of information rate for strongly non-normal networks]\label{cor:noise-cancel}
Let $B$, $C$ and $A$ be as in \cref{thm:DAD}, and assume that the pair $(S,e_q)$, with $q:=\arg\max \{d_{t_1},\dots,d_{t_p}\}$, is observable. Then,
\begin{align}
    \lim_{\frac{d_{\min}}{d_{\max}}\to 0}\cR_{T} &= -\frac{1}{\ln 2}\tr(A). \smallskip \label{eq:alpha}
\end{align}
\end{cor}
\begin{proof}
From \cref{thm:DAD}, decreasing the ratio ${d_{\min}}/{d_{\max}}$ is equivalent to decreasing the noise variance $\sigma^2$ for a network with adjacency matrix $S$. Then, \Cref{eq:alpha} directly follows from \cref{prop:asymp-SNR-low}, noting that $\tr(A)=\tr(S)$.
\end{proof}

From the above analysis, it follows that increasing the non-normality of the network by diminishing the ratio ${d_{\min}}/{d_{\max}}$ is always beneficial to the information rate $\cR_T$, in that it reduces the detrimental effect of output noise. Indeed, for the class of networks in \Cref{eq:A non-normal}, the ratio $d_{\min}/ d_{\max}$, can be thought of regulating the ``degree'' of non-normality of the network. To clarify this claim, assume that $S_{ij}\ne 0$ and consider two diagonal entries $d_{i}, d_{j}$ with $d_{i}\ne d_{j}$, in this case the $(i,j)$-th entry and $(j,i)$-th entry of $A$ take the form $\smash{A_{ij}=\frac{d_{i}}{d_{j}}S_{ij}}$ and $\smash{A_{ji}=\frac{d_{j}}{d_{i}}S_{ij}}$, respectively. As the ratio $d_{i}/d_{j}$ tends either to zero or to infinity, matrix $A$ departs from symmetry because  the entries $A_{ij}$ and $A_{ji}$ becomes very heterogenous in magnitude. As a concrete example of this, consider the chain network described in our paper (Equation 9 of the main text). The adjacency matrix of this network (here reported for convenience) reads as
\begin{align} \label{eq:chain}
    A = \begin{bmatrix}
        \gamma & \beta/\alpha & 0 & \cdots & 0 \\
         \alpha\beta & \gamma & \beta/\alpha & \ddots & \vdots \\
         0 & \alpha\beta & \ddots & \ddots & 0 \\
         \vdots & \ddots & \ddots & \ddots & \beta/\alpha \\
          0 & \cdots & 0 & \alpha\beta & \gamma \\
        \end{bmatrix}\in\mathbb{R}^{n\times n},
\end{align}
with $\beta,\alpha>0$ and $\gamma<-2\beta$. As described in our paper and in Supplementary Note 6, the parameter $\alpha$ (directionality strength) regulates the ``degree'' of non-normality of the network. Notice that the matrix in \Cref{eq:chain} can be rewritten as in \Cref{eq:A non-normal} with $D=\mathrm{diag}(1,\alpha,\dots,\alpha^{n-1})$ and 
\begin{align} \label{eq:chainD}
    S= \begin{bmatrix}
        \gamma & \beta & 0 & \cdots & 0 \\
         \beta & \gamma & \beta & \ddots & \vdots \\
         0 & \beta & \ddots & \ddots & 0 \\
         \vdots & \ddots & \ddots & \ddots & \beta \\
          0 & \cdots & 0 & \beta & \gamma \\
        \end{bmatrix}.
 \end{align}
Assume that $\alpha>1$. Then, \Cref{cor:noise-cancel} yields that, as $d_{\min}/d_{\max}=1/\alpha^{n-1}$ tends to zero, the information rate $\mathcal{R}_{T}$ tends to $-\frac{1}{\ln 2}\text{tr}(A)=-\frac{\gamma n}{\ln 2}$, which corresponds to the information rate of the network in the noiseless case. In other words, as $\alpha$ as grows unbounded (large directionality strength, and, therefore, strong non-normality), the information rate approaches the one achieved in the noiseless case.

Finally, we stress that a small value of ${d_{\min}}/{d_{\max}}$ would typically correspond to a large magnitude of some entries of $A$, which is not desirable in many scenarios. However, we also observe that, in the above example of the chain network, $d_{\min}/d_{\max}$ depends on the chain length $n$, and for large $n$ the ratio $d_{\min}/d_{\max}$ can be made small regardless of the value of $\alpha$ (provided that $\alpha>1$). Inspired by this example, we therefore propose an alternative procedure that leads to a decrease of the ratio ${d_{\min}}/{d_{\max}}$, while keeping the entries $A$ bounded in magnitude. We term this procedure ``directed stratification'' as it partitions the network in different ``directed'' layers, according to the length of the paths from a given subset of nodes to all the other network's nodes.

\begin{figure*}[h!]
\begin{center}
\includegraphics[scale=0.6]{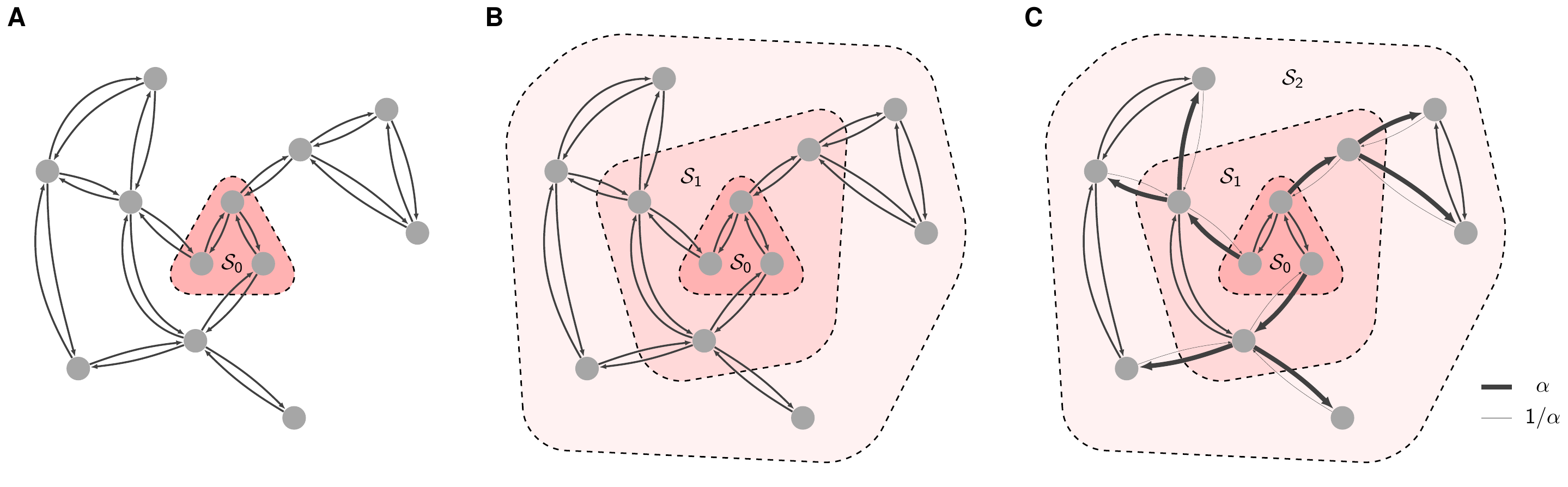} 
\end{center}
\caption{\textbf{Construction of ``layered'' non-normal networks.} Example of construction of a ``layered'' non-normal network with $n=12$ nodes. \textbf{(A)} A subset $\mathcal{S}_0$ of nodes in a weights graph $\mathcal{G}$ with homogeneous weights is selected. \textbf{(B)} Each node of the network is assigned to a different layers according to its  distance from $\mathcal{S}_0$. \textbf{(C)} The parameter $\alpha>0$ is used to generate a directional structure between adjacent~layers. The thickness of each edge in the network is proportional to their weight.}
\label{Fig:stratification}
\vspace{0.15cm}
\end{figure*}

To this end, 
given any two nodes $k,t\in\mathcal{V}$, we denote by $d(k,t)$ the length of a shortest path from the node $k$ to the node $t$. Given $\mathcal{K}\subseteq\mathcal{V}$ and $t\in\mathcal{V}$, we denote by $d(\mathcal{K},t)$ the minimum length of a shortest path from the nodes in $\mathcal{K}$ to the node $t$, namely 
\begin{align}
d(\mathcal{K},t)=\min\left\{\,d(k,t)\ :\ k\in\mathcal{K}\,\right\}.
\end{align}
Further, let $s_{\min}$ and $s_{\max}$ denote the smallest and largest (in magnitude) non-zero entries of $S$, respectively, and consider the matrix $A:=DSD^{-1}$ as in \Cref{eq:A non-normal}.
We select any subset of nodes in the network, say $\mathcal{S}_0$ , and define the following sets
\begin{align}
	\mathcal{S}_{\ell}=\left\{v\in\mathcal{V}\,:\, d(\mathcal{K},v)=\ell\right\}, \ \ell=1,2,\dots,\ell_{\max},
\end{align}
where $\ell_{\max}$ is the maximum length of a shortest path from subset $\mathcal{S}_0$ to any other node in $\mathcal{V}$. Next, we set $d_{i}=\alpha^{\ell}$, $\alpha>0$, for each node $i\in\mathcal{S}_{\ell}$. Then, for every edge $(i,j)\in\mathcal{E}$, we distinguish three cases:
\begin{enumerate}[(i)]
\item if $i,j\in\mathcal{S}_{\ell}$, then $A_{ij}=\frac{d_{j}}{d_{i}}S_{ij}=\frac{\alpha^{\ell}}{\alpha^{\ell}}S_{ij}=S_{ij}$;
\item if $i\in\mathcal{S}_{\ell-1}$ and $j\in\mathcal{S}_{\ell}$, then $A_{ij}=\frac{\alpha^{\ell-1}}{\alpha^{\ell}}S_{ij}=\frac{1}{\alpha}S_{ij}$;
\item if $i\in\mathcal{S}_{\ell}$ and $j\in\mathcal{S}_{\ell-1}$, then $A_{ij}=\frac{\alpha^{\ell}}{\alpha^{\ell-1}}S_{ij}=\alpha S_{ij}$.
\end{enumerate}
From the above equations, we observe the magnitude of the entries of $A$ in \Cref{eq:A non-normal} is bounded in the interval $[s_{\min}/\alpha,\alpha s_{\max}]$. In graphical terms, the above-described procedure corresponds to a ``directed stratification'' of $\mathcal{G}$, as illustrated in~\Cref{Fig:stratification}.

Finally, if we consider a matrix $A$ generated as above, an input matrix $B=[e_{k_1},\dots,e_{k_m}]$, such that there exists $r \in \{k_1,\dots,k_m\}$ with $r\in \mathcal{S}_0$, and an output matrix $C=[e_{t_1},\dots,e_{t_p}]^\top$, such that there exists $q \in \{k_1,\dots,k_m\}$ with $q\in \mathcal{S}_{\ell_{\max}}$, in view of \Cref{thm:DAD},
\begin{align}
    \cR_T(A,\sigma^2)\ge \cR_T\left(S,\left(\alpha^{\ell_{\max}-1} \sigma\right)^2\right).
\end{align}
The latter equation shows that it is possible to reduce the effect of noise in the rate either by increasing $\alpha$ or, for a fixed $\alpha>1$, by increasing $\ell_{\max}$. In particular, via a proper tuning of these two parameters, it is possible to increase the rate while keeping the entries of $A$ bounded in magnitude in a desired interval. Remarkably, this offers a viable way to enhance the communication performance of real-world networks. \\[0.25cm]

\section{Optimal communication architectures}

Optimal communication networks can be computed by solving the following optimization problem:\footnote{Here, we considered optimality w.r.t. the information rate $\mathcal{R}_T$. Similar considerations and results apply when considering the information capacity $\mathcal{C}_T$.}
\begin{align}
    \max_{A\in \mathcal{S}} \mathcal{R}_T(A),
\end{align}
where $\mathcal{S}$ denotes the set of real stable matrices, and we made explicit the dependence of $\mathcal{R}_T$ on $A$. This problem is non-convex and computationally expensive. In fact, it requires solving two (typically non-convex) optimization problems: one for computing the information rate, and one for finding the optimal $A$. However this problem can be made more tractable by introducing some mild assumptions and using the properties of the information rate, as we describe next.

\begin{figure}[h!]
    \centering
    \includegraphics{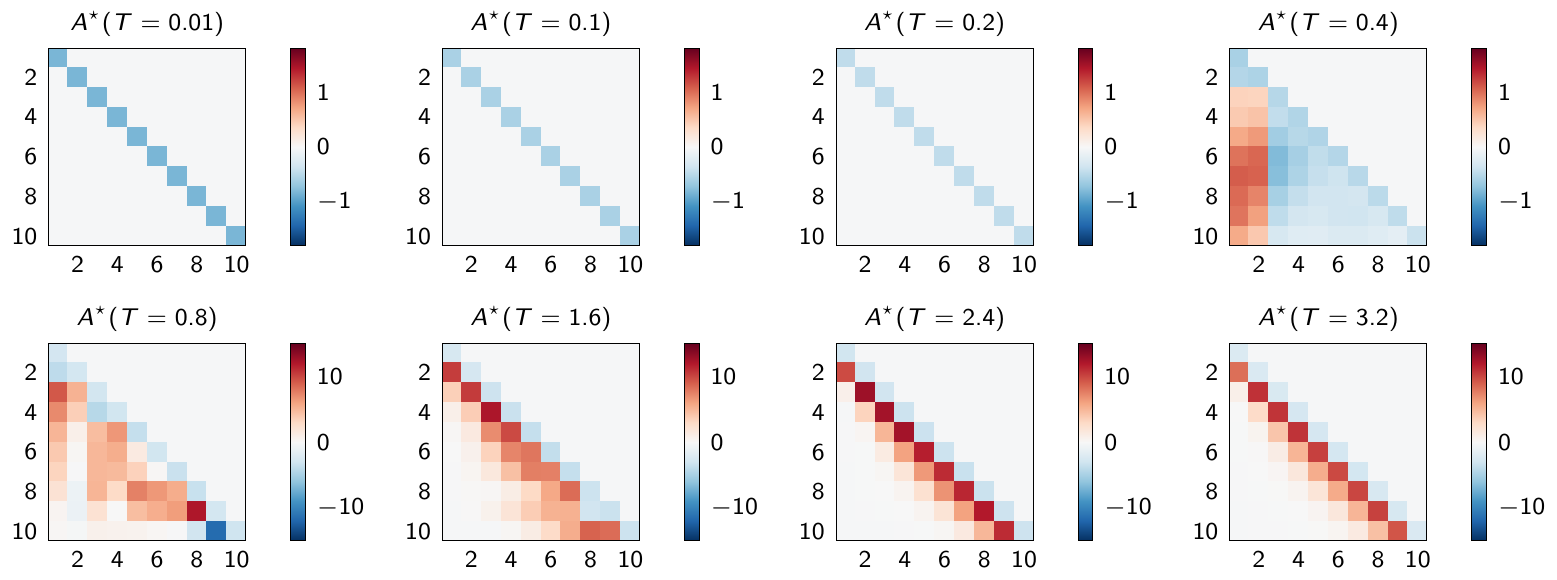}
    
    \vspace{0.5cm}
    
    \includegraphics{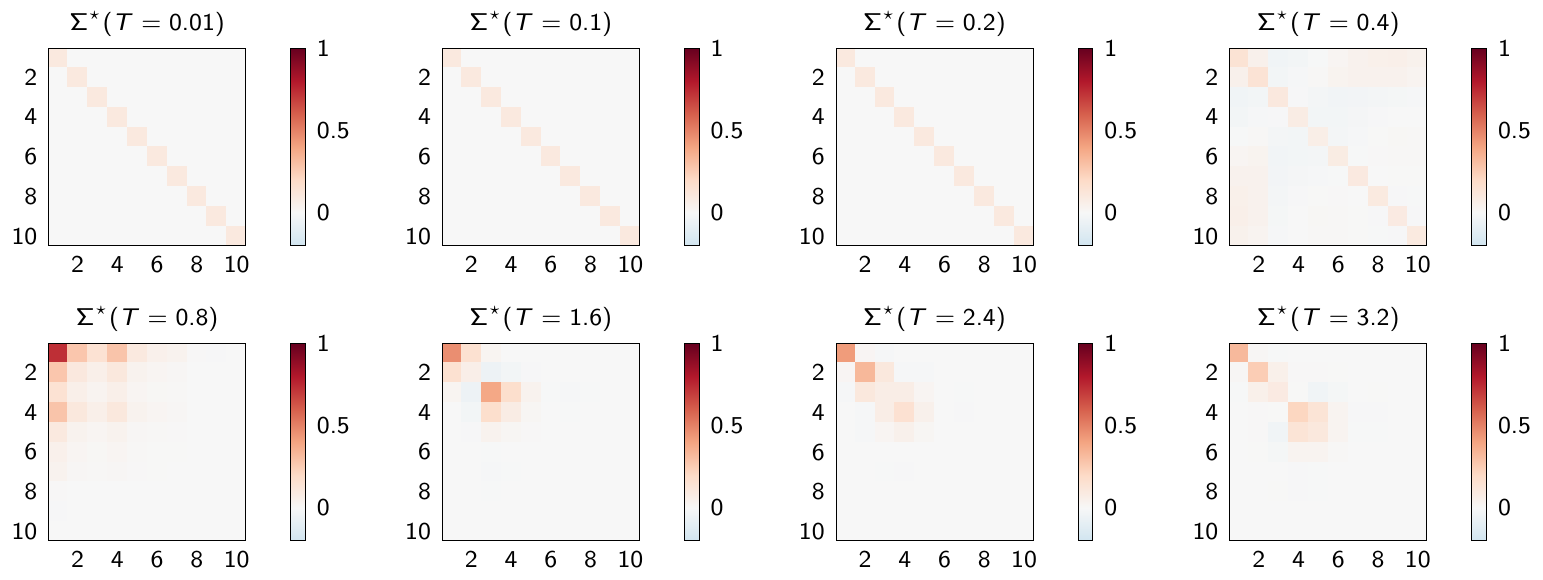}
    \caption{\textbf{Optimal networks and input covariances as a function of the transmission window $T$.} Optimal network architecture $A^\star$ and input covariance $\Sigma^\star$ obtained by solving the optimization problem in \Cref{eq:optimization2} for $n=10$, $\sigma^2=1$, $\varepsilon =2.5\times 10^{-3}$. The solution of the optimization problem  has been computed via unconstrained coordinate gradient ascent over unit-trace positive definite $\Sigma$ and lower triangular $A\in\mathcal{S}_{\Delta,r}$. The simulations have been carried out in Python using Autograd \cite{Autograd}.
    }
    \label{fig:A_opt}
\end{figure}

In what follows, we restrict the attention to the subset $\mathcal{S}_{r}\subset \mathcal{S}$ consisting of stable matrices with real eigenvalues and we consider $B=C=I$. For $C=I$, it is not difficult to see that $\mathcal{R}_T$ is invariant under unitary similarity transformation of $A$. In view of this fact, we can consider the Schur form of the matrices in $\mathcal{S}_{r}$ and reduce the original problem to an optimization problem over the set of lower triangular matrices with negative diagonal entries. We denote the latter set with $\mathcal{S}_{\Delta,r}$. The new ``simplified'' optimization problem reads as
\begin{align}\label{eq:optimization}
	\max_{A \in\mathcal{S}_{\Delta,r}}\mathcal{R}_{T}(A)-\varepsilon\|A \|_{F},
\end{align}
where we added a regularization term $\varepsilon\|A \|_{F}$, $\varepsilon>0$, to the objective function $\mathcal{R}_{T}(A)$ to bound the magnitude of the entries of $A$. Recall that the computation of the information rate $\mathcal{R}_{T}$ requires solving a constrained optimization problem over input covariances $\Sigma $. Consequently, in order to numerically solve the problem in \Cref{eq:optimization}, we exploited a coordinate gradient ascent strategy \cite[Ch.~9]{NW06} over unit-trace covariance matrices $\Sigma$,  and connectivity matrices $A \in\mathcal{S}_{\Delta,r}$. Numerical solutions of the optimization problem in \Cref{eq:optimization} are shown in \Cref{fig:A_opt}, for a network of $10$ nodes, $\sigma^2=1$, and different values of $T>0$.

From these plots, we observe that:
\begin{enumerate}
\item for small values of $T$, the optimal structure $A^{\star}$ is diagonal and, therefore, normal. In this case, the corresponding optimal input covariance $\Sigma^{\star}$ is also diagonal.
\item as $T$ increases, the strictly lower triangular entries of $A^{\star}$ becomes different from zero, yielding a non-normal optimal network structure.  In particular, for $T$ large enough, the entries of the subdiagonal of $A^{\star}$ are greater than the other lower triangular entries, resulting in an optimal structure similar to a purely feedforward chain. In this case, the non-zero entries of the corresponding optimal input covariances are localized around the upper diagonal entries (source~nodes).
\end{enumerate}

\begin{figure}[h!]
    \centering
    \includegraphics{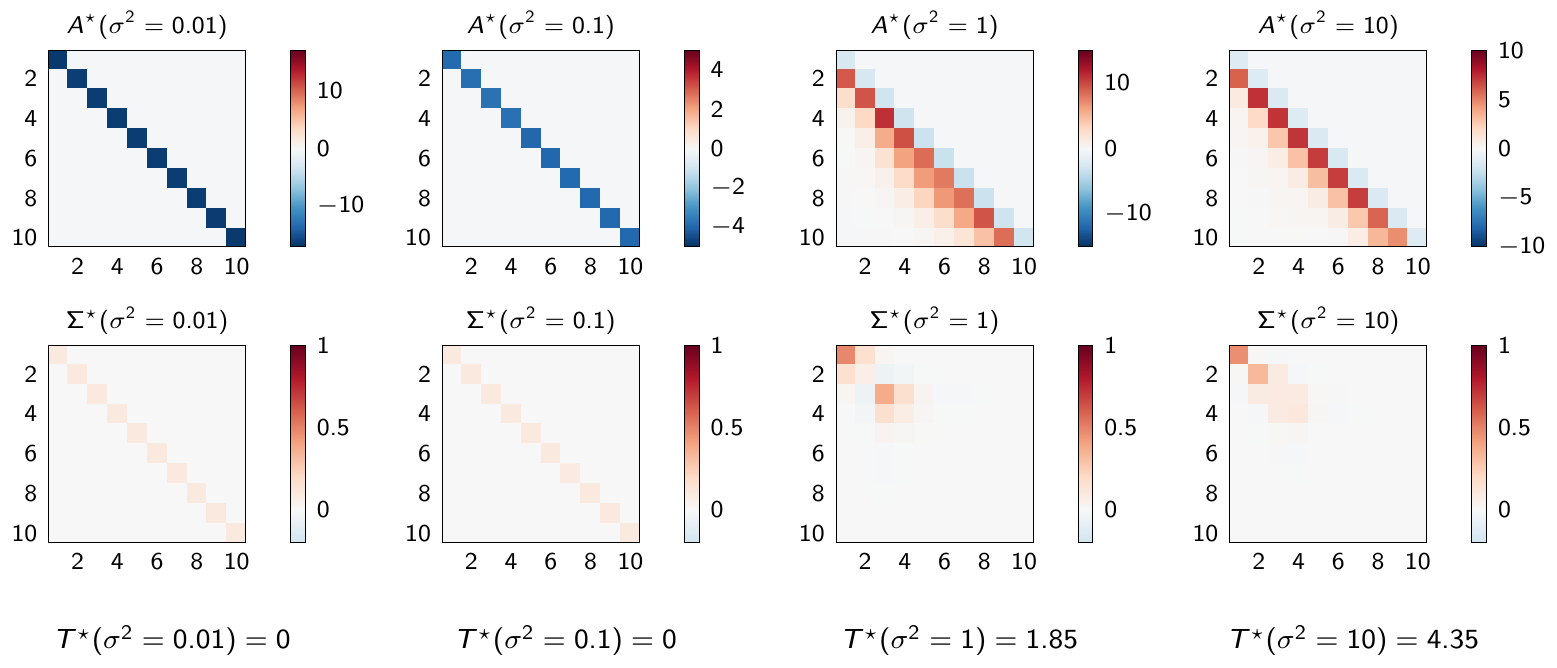}
    \caption{\textbf{Optimal networks and input covariances as a function of the noise covariance $\sigma^2$.} Optimal network architecture $A ^{\star}$, input covariance $\Sigma^{\star}$, and transmission window $T^{\star}$ obtained by solving the problem in \Cref{eq:optimization2} for $n=10$ nodes, $\varepsilon=2.5\times 10^{-3}$, and different values of $\sigma^2$. The solution of the problem in \Cref{eq:optimization2} has been computed via unconstrained coordinate gradient ascent over unit-trace positive definite $\Sigma$'s, lower triangular $A \in\mathcal{S}_{\Delta,r}$, and non-negative $T$. 
    }
    \label{fig:A_opt_noise}
\end{figure}

To investigate the effect of noise on the optimal matrix, we further optimized the cost function in \Cref{eq:optimization} over the transmission window $T\ge 0$. Namely, we considered the following regularized optimization problem:
\begin{align}\label{eq:optimization2}
	\max_{A \in\mathcal{S}_{\Delta,r},T\ge 0}\mathcal{R}_{T}(A)-\varepsilon\|A \|_{F},
\end{align}
As before, we exploited a coordinate gradient ascent method to solve the latter problem. In this case, gradient ascent need to be jointly  performed w.r.t. unit-trace covariances $\Sigma$, lower triangular connectivity matrices $A \in\mathcal{S}_{\Delta,r}$, and non-negative transmission windows $T\ge 0$. \Cref{fig:A_opt_noise} shows the numerical results obtained via this optimization strategy, for different values of $\sigma^2$. We notice that, for small values of  $\sigma^2$, the optimal transmission window equals zero, and the optimal connectivity matrix and covariance are both diagonal. For larger values of $\sigma^2$ the optimal transmission window becomes strictly positive, and grows as $\sigma^2$ increases. Further, the resulting $A^\star$ is non-normal and resembles a feedforward chain, while $\Sigma^{\star}$ has non-zero diagonal entries concentrated around the upper diagonal entries.

\clearpage

\newsavebox\mytempbib
\savebox\mytempbib{\parbox{\textwidth}{
}}
